\documentclass[preprint2,longabstract]{aastex}

\usepackage{url}
\usepackage{subfigure}
\usepackage{longtable}
\usepackage{lscape}
\usepackage{multirow}
\usepackage{color}
\usepackage{textcomp}
\usepackage{natbib}
\usepackage[colorlinks,linkcolor=red,anchorcolor=green,citecolor=blue]{hyperref}

\newcommand{\HII}{H {\small{II}} }
\newcommand{\kms}{{\rm km\,s}^{-1}}
\newcommand{\Msun} {M_{\sun}}

\newcommand{\mjyb}{{\rm mJy~beam}^{-1}}

\slugcomment{}

\shorttitle{The TOP-SCOPE survey of PGCCs}
\shortauthors{Zhang et al.}

\begin{document}

\title{The TOP-SCOPE survey of PGCCs: PMO and SCUBA-2 observations of 64 PGCCs in the 2nd Galactic Quadrant}

\author{Chuan-Peng Zhang\altaffilmark{1,2,*},
Tie Liu\altaffilmark{3,4},
Jinghua Yuan\altaffilmark{1},
Patricio Sanhueza\altaffilmark{5},
Alessio Traficante\altaffilmark{6},
Guang-Xing Li\altaffilmark{7},
Di Li\altaffilmark{1},
Ken'ichi Tatematsu\altaffilmark{5},
Ke Wang\altaffilmark{8},
Chang Won Lee\altaffilmark{9,10},
Manash R. Samal\altaffilmark{11},
David Eden\altaffilmark{12},
Anthony Marston\altaffilmark{13},
Xiao-Lan Liu\altaffilmark{1},
Jian-Jun Zhou\altaffilmark{14},
Pak Shing Li\altaffilmark{15},
Patrick M. Koch\altaffilmark{16},
Jin-Long Xu\altaffilmark{1},
Yuefang Wu\altaffilmark{17},
Mika Juvela\altaffilmark{18},
Tianwei Zhang\altaffilmark{17},
Dana Alina\altaffilmark{19},
Paul F. Goldsmith\altaffilmark{20},
L. V. T\'oth\altaffilmark{21},
Jun-Jie Wang\altaffilmark{1},
Kee-Tae Kim\altaffilmark{3}
}


\altaffiltext{*}{Email: cpzhang@nao.cas.cn}
\altaffiltext{1}{National Astronomical Observatories, Chinese Academy of Sciences, 100012 Beijing, PR China}
\altaffiltext{2}{Max-Planck-Institut f\"ur Astronomie, K\"onigstuhl 17, 69117 Heidelberg, Germany}
\altaffiltext{3}{Korea Astronomy and Space Science Institute 776, Daedeokdae-ro, Yuseong-gu, Daejeon 34055, Korea}
\altaffiltext{4}{East Asian Observatory, 660 N. A'oh$\rm \bar{o}k\bar{u}$ Place, Hilo, Hawaii 96720-2700, USA}
\altaffiltext{5}{National Astronomical Observatory of Japan, National Institutes of Natural Sciences, 2-21-1 Osawa, Mitaka, Tokyo 181-8588, Japan}
\altaffiltext{6}{IAPS - INAF, via Fosso del Cavaliere, 100, I-00133 Roma, Italy}
\altaffiltext{7}{University Observatory Munich, Scheinerstrasse 1, D-81679 Munich, Germany}
\altaffiltext{8}{European Southern Observatory, Karl-Schwarzschild-Str.2, D-85748 Garching bei M\"unchen, Germany}
\altaffiltext{9}{Korea Astronomy \& Space Science Institute (KASI), 776 Daedeokdae-ro, Yuseong-gu, Daejeon 305-348, Republic of Korea}
\altaffiltext{10}{University of Science \& Technology, 176 Gajeong-dong, Yuseong-gu, Daejeon, Republic of Korea}
\altaffiltext{11}{Graduate Institute of Astronomy, National Central University 300, Jhongli City, Taoyuan County 32001, Taiwan}
\altaffiltext{12}{Astrophysics Research Institute, Liverpool John Moores University, IC2, Liverpool Science Park, 146 Brownlow Hill, Liverpool, L3 5RF, UK}
\altaffiltext{13}{ESA/STScI, 3700 San Martin Dr, Baltimore, MD 21218, USA}
\altaffiltext{14}{Xinjiang Astronomical Observatory, CAS, 150, Science 1-street, 830011 Urumqi, PR China}
\altaffiltext{15}{Astronomy Department, University of California, Berkeley, CA 94720, USA}
\altaffiltext{16}{Academia Sinica, Institute of Astronomy and Astrophysics, P.O. Box 23-141, Taipei 106, Taiwan}
\altaffiltext{17}{Department of Astronomy, Peking University, 100871 Beijing, PR China}
\altaffiltext{18}{Department of Physics, P.O.Box 64, FI-00014, University of Helsinki, Finland}
\altaffiltext{19}{Physics Department, Nazarbayev University, Kabanbay batyr avenue 53, 010000 Astana, Kazakhstan}
\altaffiltext{20}{Jet Propulsion Laboratory, California Institute of Technology, 4800 Oak Grove Drive, Pasadena, CA 91109, USA}
\altaffiltext{21}{Lor\'and E\"otv\"os University, Department of Astronomy, P\'azm\'any P.s. 1/a, H-1117 Budapest, Hungary}

\begin{abstract}

In order to understand the initial conditions and early evolution of star formation in a wide range of Galactic environments, we carried out an investigation of 64 \textit{Planck} Galactic Cold Clumps (PGCCs) in the second quadrant of the Milky Way. Using the $^{13}$CO and C$^{18}$O $J = 1 - 0$ lines, and 850\,$\mu$m continuum observations, we investigated cloud fragmentation and evolution associated with star formation. We extracted 468 clumps and 117 cores from the $^{13}$CO line and 850\,$\mu$m continuum maps, respectively. We make use of the Bayesian Distance Calculator and derived the distances of all 64 PGCCs. We found that in general, the mass-size plane follows a relation of $m\sim r^{1.67}$.  At a given scale, the masses of our objects are around 1/10 of that of typical Galactic massive star-forming regions. Analysis of the clump and core masses, virial parameters, densities, and mass-size relation suggests that the PGCCs in our sample have a low core formation efficiency ($\sim$3.0\%), and most PGCCs are likely low-mass star-forming candidates. Statistical study indicates that the 850\,$\mu$m cores are more turbulent, more optically thick, and denser than the $^{13}$CO clumps for star formation candidates, suggesting that the 850\,$\mu$m cores are likely more appropriate future star-formation candidates than the $^{13}$CO clumps.

\end{abstract}



\section{Introduction}

Stars form in the dense, cold regions within molecular clouds. However, the physical and chemical properties of the cold compact objects that breed stars are still poorly understood. Stars could form  out of  gravitationally bound substructures within a molecular cloud, but how the substructures themselves form is strongly debated \citep[e.g.,][]{Johnstone2004}. Investigating the cloud fragmentation from large scale to small scale may be one way to determine this. An important approach to improve our understanding is to perform a statistical study towards the cold dense clumps from unbiased large surveys in the Milky Way.

Fortunately, the \textit{Planck} satellite has allowed for a systematically extracted inventory of Galactic cold clumps \citep{Planck2011a1} using multiple bands from submillimeter to millimeter wavelengths. The Cold Core Catalogue of \textit{Planck} Objects (C3PO) consisting of 10,783 cold cores \citep{Planck2011a23}, and the \textit{Planck} Early Release Cold Cores Catalog (ECC), the sub-catalog, containing 915 of the most reliable detections were released in 2011 \citep{Planck2011a7}. The C3PO was the first unbiased, all-sky catalogue of cold objects, and gives an unprecedented statistical view to the properties of these potential pre-stellar clumps and offers a unique possibility for their classification in terms of their intrinsic properties and environment \citep{Planck2011a23}. The cores in C3PO have relatively high column densities (0.1 $\sim$ 1.6 $\times$ 10$^{22}$ cm$^{-2}$) and low dust temperatures \citep[$\sim$10 -- 15 K,][]{Planck2011a22,Planck2011a23}. This was followed by the \textit{Planck} Catalogue of Galactic Cold Clumps (PGCCs; \citealt{Planck2016a28}), an all-sky catalogue of Galactic cold clump candidates, containing 13,188 Galactic sources, detected by \textit{Planck}. This catalogue is the full version of the ECC catalogue. The \textit{Herschel} key programme ``Galactic Cold Cores" was a follow-up to study the substructure and physics of selected C3PO sources (selection being performed on their intrinsic properties and their Galactic location). This study commenced during the \textit{Herschel} Science Demonstration Phase Data \citep{Juvela2010}.

Further follow-up studies of PGCC objects have been carried out with ground-based telescopes to study the evolutionary conditions of PGCCs. These facilities include: the James Clerk Maxwell Telescope (JCMT), the Purple Mountain Observatory (PMO), the Nobeyama Radio Observatory, the Taeduk Radio Astronomy Observatory (TRAO), the Korean VLBI Network (KVN), the Caltech Submillimeter Observatory (CSO), the Submillimeter Array (SMA), and the Institut de radioastronomie millim\'{e}trique (IRAM) \citep{Wuyf2012,Meng2013,Liu2012,Liu2013,Liu2015,Liu2016,Yuan2016,Zhangtw2016,Kim2017,Tatematsu2017,Tang2018}. These ground-based studies allow us to improve our understanding of dense cores and star formation in widely different environments at higher spatial resolution than the \textit{Planck} observations, using different tracers from the continuum to spectral lines (e.g., CO, N$_2$H$^+$, HCO$^+$). For example, \citet{Wuyf2012} and \citet{Meng2013} carried out a survey towards 745 PGCCs in $^{12}$CO, $^{13}$CO, C$^{18}$O $J = 1-0$ using the PMO 13.7-m telescope. They found a variety of morphologies from extended diffuse to dense, isolated, cometary, and filamentary structures. They also found that the PGCCs are the most quiescent among the sample of weak-red IRAS, infrared dark clouds, UC \HII candidates, extended green objects, and methanol maser sources. \citet{Liu2016} performed a series of observations with ground-based telescopes towards one PGCC in the $\lambda$ Orionis complex to systematically investigate the effects of stellar feedback. Particularly they discovered an extremely young Class 0 protostellar object (G192N) and a proto-brown dwarf candidate (G192S), located in a gravitationally bound bright-rimmed clump. This provides a sample to study the earliest stage of star formation. \citet{Yuan2016} conducted the first large survey of dense gas toward PGCCs in the $J = 1-0$ transitions of HCO$^+$ and HCN toward 621 molecular cores. On the basis of an inspection of the derived density information given in their PGCC catalog, \citet{Yuan2016} suggested that about 1000 out of 13,188 PGCCs show a sufficient reservoir of dense gas to form stars.

Based on the studies mentioned above, PGCCs are cold ($\sim$ 10-15\,K), turbulence-dominated, and have relatively low column densities compared to other star-forming regions \citep{Wuyf2012,Planck2011a22,Planck2011a23}. Additionally, most clumps are quiescent and lack signs of star formation, indicating that the PGCCs are most likely in the very initial evolutionary stages of star formation \citep{Wuyf2012,Yuan2016}. Furthermore, previous studies indicate that gaseous CO abundance (or depletion) can be used as a tracer for the evolution of molecular clouds \citep{Liu2013,Zhangcp2017}.

The work described here is part of the TOP-SCOPE\footnote{TOP: TRAO observations of \textit{Planck} cold clumps; SCOPE: SCUBA-2 Continuum Observations of Pre-protostellar Evolution} survey of PGCCs, which combines the TRAO 13.7-m telescope and the Submillimetre Common-User Bolometer Array 2 (SCUBA-2; \citealt{Holland2013}) instrument on board of the JCMT to observe around 1000 PGCCs \citep[\textcolor{blue}{Eden et al., in prep.};][]{Liu2017}. It is a follow-up study of \citet{Zhangtw2016}, who mainly used $^{12}$CO and $^{13}$CO $J=1-0$ emission lines to investigate the gas content of 96 PGCCs from $98^\circ < l < 180^\circ$ and $-4^\circ < b < 10^\circ$ in the second quadrant of the Milky Way. The survey has covered most of the densest ECCs in the regions of the 2nd quadrant \citep[see more details in][]{Zhangtw2016}. \citet{Zhangtw2016} discussed the properties and morphologies of these clumps combining the distributions of excitation temperature, velocity dispersion, and column density. The second quadrant is home to many well-known star formation regions, such as W3, W4, W5, NGC\,7129, NGC\,7538, and S235 \citep{dame1987,dame2001,Heyer1998}. A systematic cold core analysis of the second quadrant could thus be essential for understanding the properties of the initial star-forming conditions in the Outer Galaxy.

\begin{figure}[h]
\centering
\includegraphics[width=0.45\textwidth, angle=0]{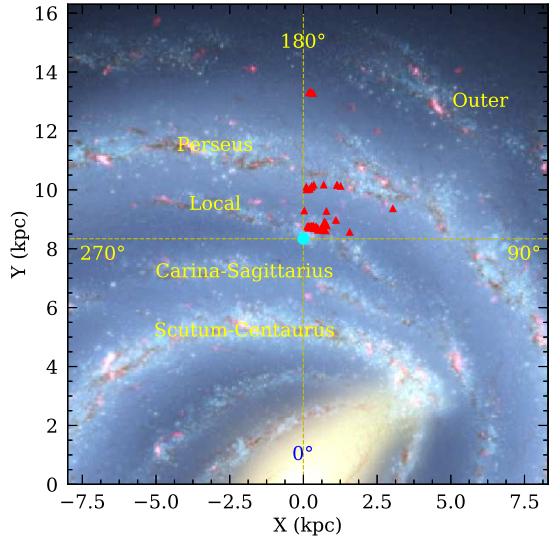}
\caption{Distribution of the clumps (red-filled triangles) on the background of an artist’s conception of the Milky Way (R. Hurt: NASA/JPL-Caltech/SSC). All sources are located in the second quadrant of the Galaxy.}
\label{Fig:mw_face}
\end{figure}

In the 96 \citet{Zhangtw2016} PGCCs, there are 64 sources that have been covered by both the SCUBA-2 850\,$\mu$m continuum and PMO $^{13}$CO, C$^{18}$O $J=1-0$ line observations. In this work, we study the 64 PGCCs mainly combining the continuum and line data mentioned above. The $^{13}$CO are C$^{18}$O are more suitable tracers to study dense conditions of the PGCCs than the $^{12}$CO and $^{13}$CO investigation in \citet{Zhangtw2016}.  These data are also compared with the WISE 12 and 22\,$\mu$m emission. The full sample is presented in Table\,\ref{Tab:64_clumps} and Fig.\,\ref{Fig:mw_face}. Section\,\ref{sect:observation} presents the observations and data reduction. Section\,\ref{sect:results} shows the results of observations and the data analysis. In Section\,\ref{sect:discussion}, we discuss the fragmentation and evolution associated with star formation, and present a statistical analysis of the morphology, velocity dispersion, virial parameter, surface density, optical depth, and excitation temperature for the $^{13}$CO clumps and 850\,$\mu$m cores. Finally, a summary is presented in Section\,\ref{sect:summary}.

\begin{figure*}
\centering
\includegraphics[width=0.32\textwidth, angle=0]{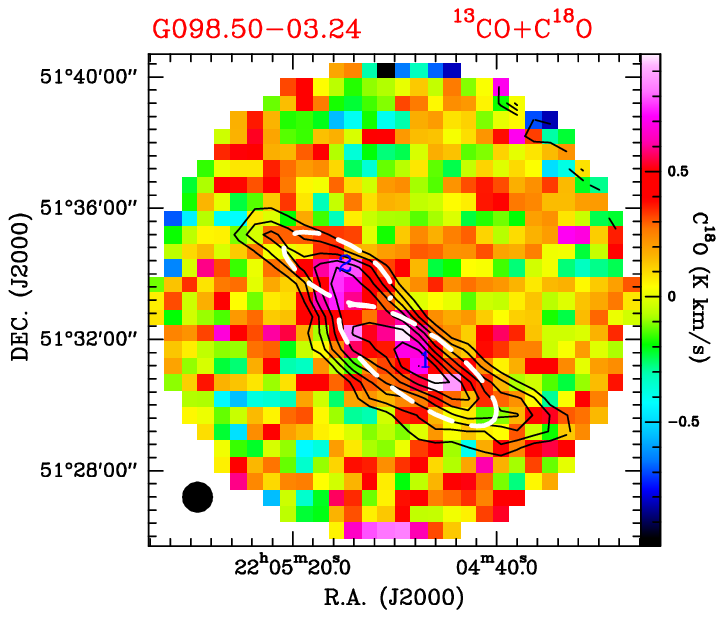}
\includegraphics[width=0.32\textwidth, angle=0]{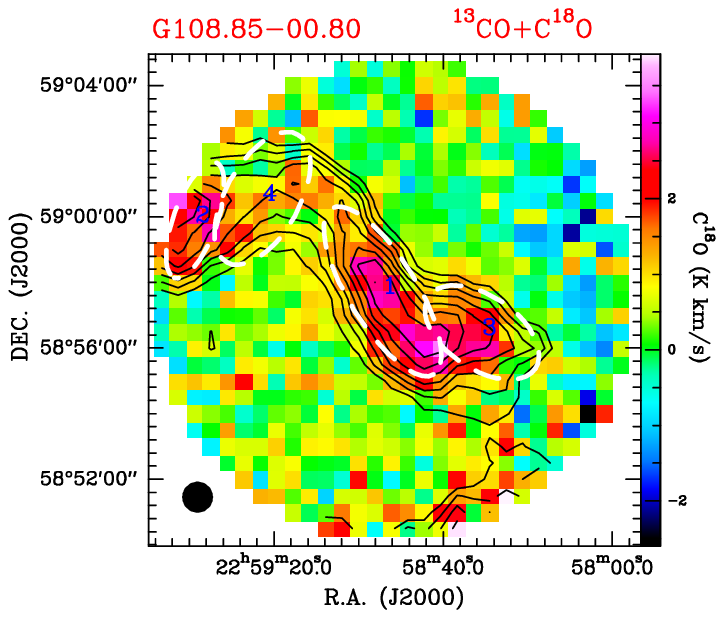}
\includegraphics[width=0.32\textwidth, angle=0]{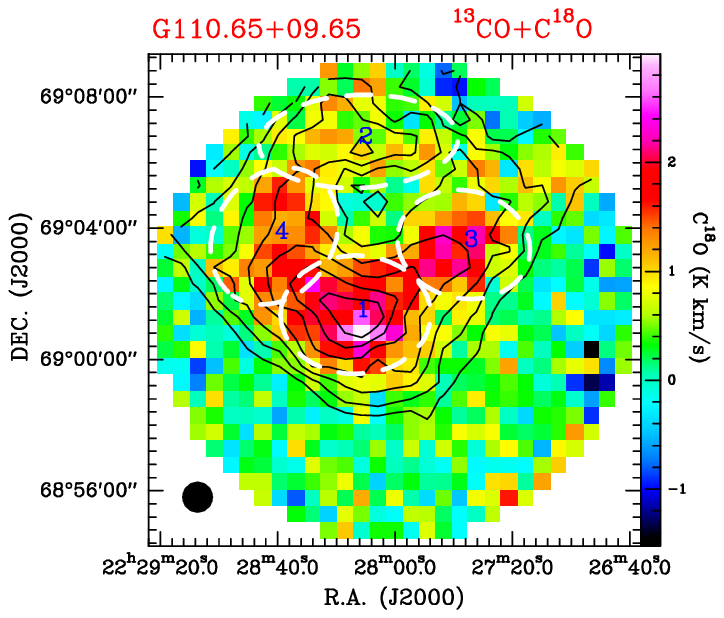}
\caption{Integrated-intensity maps of C$^{18}$O emission of each source with overlaid contours of the $^{13}$CO line. The integrated velocity ranges used are indicated within the red window in the corresponding spectrum of Fig.\,\ref{Fig:c18o-spectra}. The contour levels of the $^{13}$CO lines are drawn at 10\% steps, starting with 30\% of the peak value. The white ellipses indicate the extracted $^{13}$CO clumps. The beam size of the $^{13}$CO data is indicated in the bottom-left corner. \texttt{Supplementary figures can be downloaded in https://zcp521.github.io/pub/Figs.zip.}}
\label{Fig:13co-c18o}
\end{figure*}

\section{Observations}
\label{sect:observation}

\subsection{The CO data of the PMO 13.7-m telescope}

The CO observations were made during April -- May 2011, and December 2011 -- January 2012 using the 13.7-m millimeter telescope of Qinghai Station at the PMO\footnote{http://www.dlh.pmo.cas.cn/}. The 9-beam SIS superconducting receiver with beams separated by around 180$''$, was used as the front end. The receiver was operated in the sideband separation of single sideband mode, allowing for simultaneous observations of three CO $J = 1 - 0$ isotopologues, with $^{12}$CO in the upper sideband (USB) and $^{13}$CO and C$^{18}$O in the lower sideband (LSB). The half-power beam width (HPBW) is $52''\pm3''$, with a main beam efficiency of $\sim$\,50\% for $^{13}$CO and C$^{18}$O observations. The $^{13}$CO and C$^{18}$O data are used here. The pointing and tracking accuracies are better than 5$''$. The typical system temperatures during the runs are $\sim$\,120 K at 110.2 GHz, and varied by $\sim$\,10\% between beams. A fast Fourier transform (FFT) spectrometer was used as the back end with a total bandwidth of 1 GHz and 16,384 channels, giving a velocity resolution of $\sim$\,0.16\,$\kms$ for the $^{13}$CO and C$^{18}$O lines.

\begin{figure*}
\centering
\includegraphics[width=0.32\textwidth, angle=0]{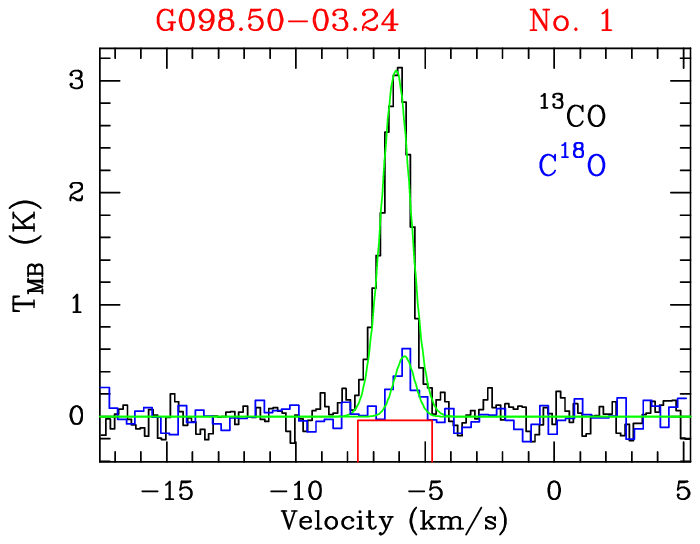}
\includegraphics[width=0.32\textwidth, angle=0]{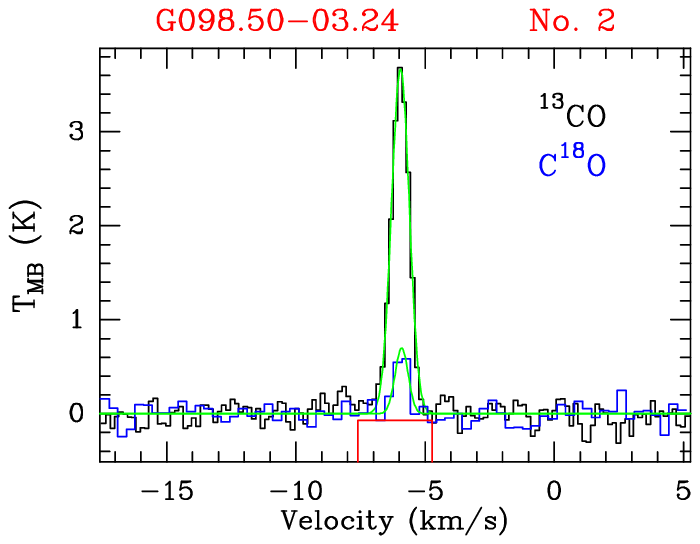}
\includegraphics[width=0.32\textwidth, angle=0]{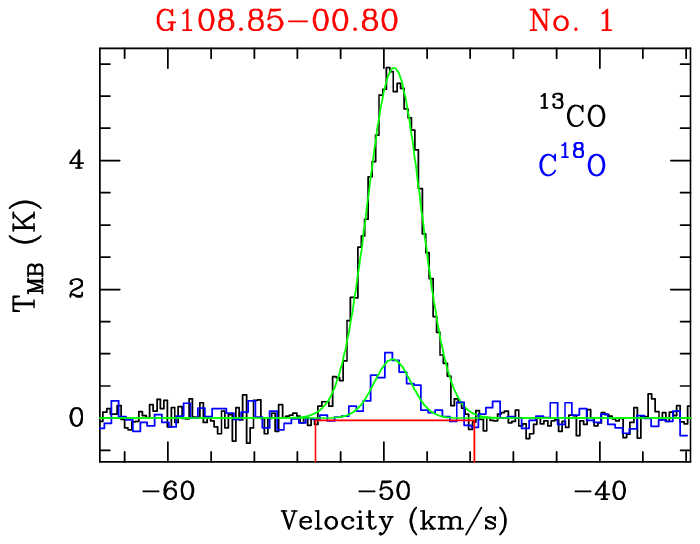}
\caption{Averaged $^{13}$CO (black line) and C$^{18}$O (blue lines) lines within the size of each extracted $^{13}$CO clump (see Fig.\,\ref{Fig:13co-c18o}). The green lines show the Gaussian fits in each spectrum. The red window indicates the velocity range corresponding to the $^{13}$CO and C$^{18}$O integrated intensity maps (see Fig.\,\ref{Fig:13co-c18o}). \texttt{Supplementary figures can be downloaded in https://zcp521.github.io/pub/Figs.zip.}}
\label{Fig:c18o-spectra}
\end{figure*}

\begin{figure*}
\centering
\includegraphics[width=0.32\textwidth, angle=0]{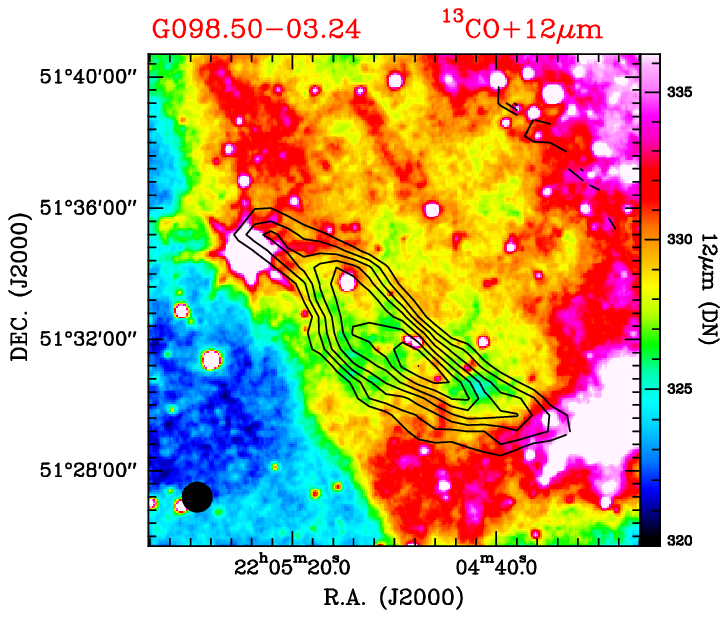}
\includegraphics[width=0.32\textwidth, angle=0]{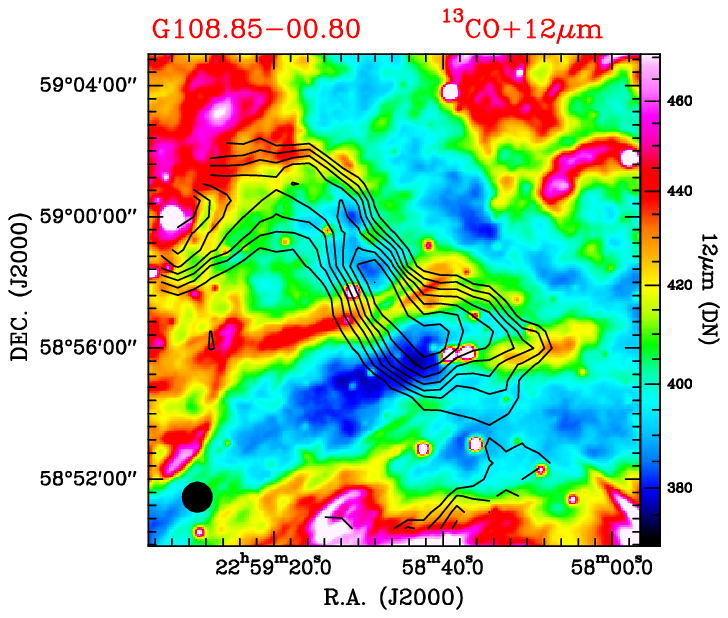}
\includegraphics[width=0.32\textwidth, angle=0]{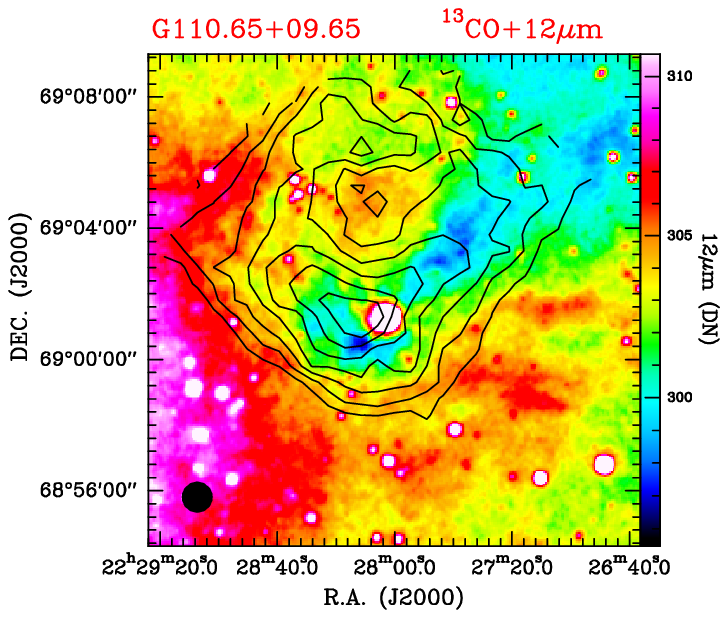}
\caption{WISE 12\,$\mu$m emission for each PGCC with overlaid $^{13}$CO contours. The contour levels of the $^{13}$CO lines are drawn at 10\% steps, starting with 30\% of the peak value. The beam size of $^{13}$CO data is indicated in the bottom-left corner. \texttt{Supplementary figures can be downloaded in https://zcp521.github.io/pub/Figs.zip.}}
\label{Fig:13co-12um}
\end{figure*}

An on-the-fly (OTF) observing mode was used for the mapping observations at a scan speed of 50$\arcsec$~s$^{-1}$. The off position for each ``off'' source was carefully chosen from an area within a 3$^{\circ}$ radius of each ``on'' source, where there is extremely weak or no CO emission \citep{dame1987,dame2001}. The antenna continuously scanned a region of 22$\arcmin\times22\arcmin$ centered on each clump, while only the central 14$\arcmin\times14\arcmin$ region is used due to the noisy edges of the OTF maps. The rms noise level was 0.1\,K in the main beam antenna temperature $T{^{*}_{\rm A}}$ for $^{13}$CO and C$^{18}$O $J = 1 - 0$. The OTF data were resampled in a three-dimensional (3D) cube with a grid spacing of 30$\arcsec$. The IRAM software package GILDAS\footnote{http://iram.fr/IRAMFR/GILDAS/} was used for the data reduction. The reduced images are presented in Figs.\,\ref{Fig:13co-c18o} and \ref{Fig:c18o-spectra}. The integrated-intensity maps of $^{13}$CO line are also overlaid on the WISE 12 and 22\,$\mu$m emission maps in Figs.\,\ref{Fig:13co-12um} and \ref{Fig:13co-22um}.

\subsection{The 850\,$\mu$m data of the JCMT 15-m telescope}

The majority of the SCUBA-2 850\,$\mu$m observations were conducted as part of the SCOPE project \citep{Liu2017}. The rest of the data were collected from the JCMT data archive of the Canadian Astronomy Data Centre (CADC). SCUBA-2 is a bolometer detector at the JCMT 15-m telescope with $\sim$10,000 pixels over eight science arrays which simultaneously observe 450 and 850\,$\mu$m with a field-of-view of 8$'$, and the effective beam size is around 10$''$ at 450\,$\mu$m and 14$''$ at 850\,$\mu$m \citep{Holland2013}. The observations used the constant velocity (CV) Daisy mode \citep{Bintley2014}, which is more sensitive in the central 3$'$ radii, and designed for small and compact sources. The 225 GHz opacity during the observations was in the range of 0.09 to 0.11, therefore we only use the 850\,$\mu$m data as the 450\,$\mu$m data isn't photometric. The data were reduced using SMURF in the STARLINK package \citep{Chapin2013,Dempsey2013}. The mapped areas were about 12$\arcmin\times12\arcmin$. The rms level in the central 3$'$ area of the maps was typically 6 -- 10 $\mjyb$. The images are presented in Fig.\,\ref{Fig:13co-850um}.

SCUBA-2 continuum observations at 850\,$\mu$m are known to be affected by contamination from spectral lines \citep{Johnstone2003,Parsons2017}, especially the $^{12}$CO $J=3-2$ line at 345.796\,GHz \citep{Drabek2012}. A typical level of the CO contamination is $<$\,20\% \citep{Nutter2007,Buckle2015,Rumble2015,Moore2015}, which is not significant. In a study of 90 PGCCs, \citet{Juvela2017} found that the CO contamination levels in SCUBA-2 images are $\leqslant$5\%. Therefore we don't correct for it here.

The observatory produced Flux Conversion Factor (FCF) was calculated using a 60$''$ aperture. If we have a clump that is bigger or smaller than this nominal FCF, we will need to adjust the flux values accordingly. The integrated flux density $S_{\rm 850\,{\mu}m}$ of each extracted core (see Section\,\ref{sect_extraction}) has been corrected using aperture correction factor provided by \citet{Dempsey2013}.

\subsection{Archival WISE data}
\label{sect:obs_wise}

NASA's Wide-field Infrared Survey Explorer \citep[WISE;][]{Wright2010} mapped the sky at 3.4, 4.6, 12, and 22\,$\mu$m (W1, W2, W3, and W4) with an angular resolution of 6.1$''$, 6.4$''$, 6.5$''$, and 12.0$''$ in the four bands, respectively. WISE achieved 5$\sigma$ point source sensitivities better than 0.08, 0.11, 1, and 6\,mJy in unconfused regions on the ecliptic in the four bands. The sensitivity was better toward the ecliptic poles due to denser coverage and lower zodiacal background. In this work, WISE 12 and 22\,$\mu$m image data are used. Additionally, the AllWISE Data in VizieR Online Data Catalog \citep{Cutri2013,Cutri2014} are used for point source cross identification (within 10$''$ radii of the peak position of each 850\,$\mu$m core) with our 850\,$\mu$m catalog (using \textit{Gaussclumps} procedure; see Section\,\ref{sect_extraction}) listed in Table\,\ref{Tab:850_clumps_2}. The WISE images are presented in Figs.\,\ref{Fig:13co-12um} and \ref{Fig:13co-22um}.

\section{Results and Analysis}
\label{sect:results}

\subsection{Distance Determination}

The distances to the PGCCs are estimated using the Bayesian Distance Calculator\footnote{http://bessel.vlbi-astrometry.org/bayesian} \citep{Reid2016}, which uses trigonometric parallaxes from the BeSSeL (Bar and Spiral Structure Legacy Survey\footnote{http://bessel.vlbi-astrometry.org/home}) and VERA (Japanese VLBI Exploration of Radio Astrometry\footnote{http://veraserver.mtk.nao.ac.jp/}) projects, to significantly improve the accuracy and reliability of kinematic distance estimates to other sources that are known to follow the Milky Way spiral structure. Based on the $^{13}$CO centroid velocity of each 850\,$\mu$m core No.\,1 within each PGCC (see Table\,\ref{Tab:850_clumps_1}), the corresponding distance parameters (distributed between 0.42 and 5.0\,kpc) are derived and listed in Table\,\ref{Tab:64_clumps}. The probabilities of the adopted distances are also listed in Table\,\ref{Tab:64_clumps}. In Fig.\,\ref{Fig:mw_face}, we present the distribution of the PGCCs on an artist's conception of the Milky Way \citep{Yuan2017}. We find that most PGCCs are located in the Local and Perseus arms with a significant population in the corresponding interarm region, while only four PGCCs (G176.17-02.10, G177.09+02.85, G177.14-01.21, and G177.86+01.04) are located in the Outer arm. We note that some derived distances are different from those in \citet{Zhangtw2016}, who used only the Galactic rotation curve to acquire the kinematic distances (distribution between 0.1 and 28.7 kpc) following the method of \citet{Sofue2011}.

\begin{figure*}
\centering
\includegraphics[width=0.32\textwidth, angle=0]{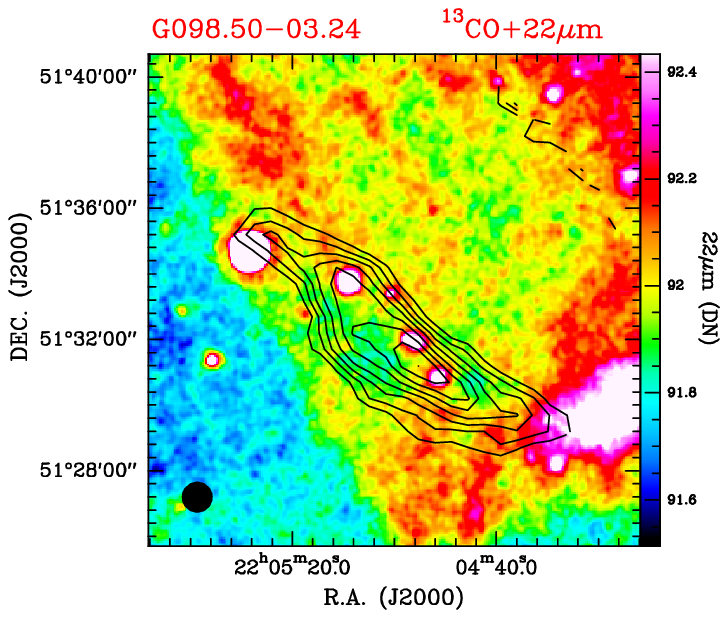}
\includegraphics[width=0.32\textwidth, angle=0]{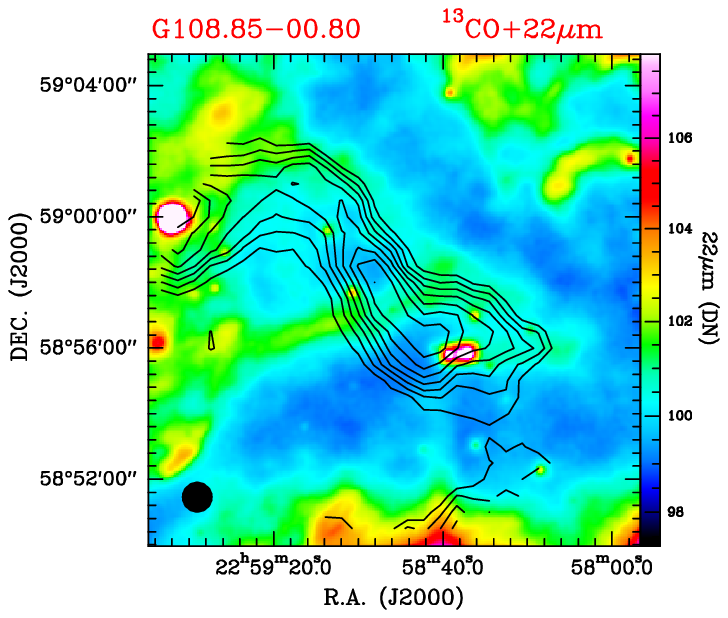}
\includegraphics[width=0.32\textwidth, angle=0]{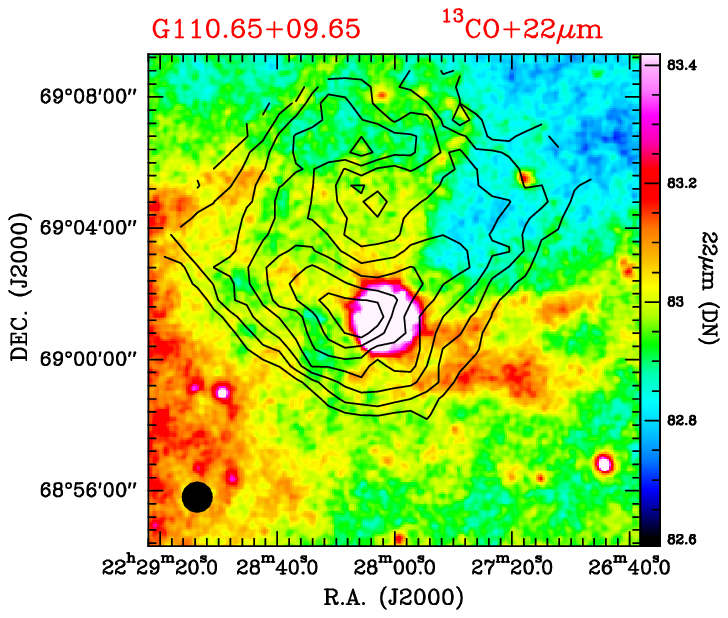}
\caption{WISE 22\,$\mu$m emission for each PGCC with overlaid $^{13}$CO contours. The contour levels of the $^{13}$CO lines are drawn at 10\% steps, starting with 30\% of the peak value. The beam size of $^{13}$CO data is indicated in the bottom-left corner. \texttt{Supplementary figures can be downloaded in https://zcp521.github.io/pub/Figs.zip.}}
\label{Fig:13co-22um}
\end{figure*}

\subsection{Fragment extraction and definition}
\label{sect_extraction}

The potential cloud fragments are extracted from the $^{13}$CO integrated line intensity and 850\,$\mu$m continuum maps with the  \textit{Gaussclumps} procedure \citep{Kramer1998,Stutzki1990,Zhang2017} in the GILDAS software package. \textit{Gaussclumps} fits a 2-dimensional fragment to the local maximum of the input cube, subtracts this fragment from the cube, creating a residual map, and then continues with the maximum of this residual map \citep{Gomez2014}. This procedure is then repeated until a stop criterion is met. We only consider fragments with peak $^{13}$CO and 850\,$\mu$m intensities of above 5$\sigma$ with the initial FWHM set at 1.1 times the beam size. The initial aperture FWHM and aperture cutoff are set as 2.0 and 8.0 times the beam size, respectively \citep[see also detailed example of configurations in][]{Belloche2011}. Considering that some extracted sources are in filamentary structures, we have rejected sources with aspect ratios larger than 5 as the study of filaments in the SCOPE PGCCs will be the subject of a further study. The measured parameters are listed in Tables\,\ref{Tab:co_clumps_1}, \ref{Tab:co_clumps_2}, \ref{Tab:850_clumps_1}, and \ref{Tab:850_clumps_2}, and indicated with ellipses in Figs.\,\ref{Fig:13co-c18o} and \ref{Fig:850um-clumps}.

In this work, we adopt ``fragment'' as the general name for both extracted clumps and cores. We consider a clump to have a typical size of 0.3 -- 3\,pc with a mass of 50 -- 500\,$\Msun$ and cores are an order of magnitude lower with sizes of 0.03 -- 0.2\,pc with masses of 0.5 -- 5\,$\Msun$ \citep[e.g.,][]{Bergin2007,Motte2017}. Based on the effective radii in Tables \ref{Tab:co_clumps_1} and \ref{Tab:850_clumps_1}, we thus refer to the $^{13}$CO objects as clumps and the 850\,$\mu$m objects as cores. Massive clouds tend to fragment into clusters of clumps and cores \citep{Pokhrel2017}, in which young stars form. Therefore, we can explore the habitats of clumps at larger scales and the cores at smaller scales, studying the fragmentation process\footnote{A caveat here is that the clumps and cores could just be discrete self-gravitating structures in a large-scale cloud (see also Section\,\ref{sect:virial}).}.

\subsection{$^{13}$CO clumps}

In total, we have extracted 468 $^{13}$CO clumps having an effective radius range of 0.1 -- 3.3\,pc with a median value of 0.4\,pc and a detected mass range of 1 -- 6132\,$\Msun$ with a median value of 66\,$\Msun$ for the clumps. Figure\,\ref{Fig:13co-c18o} shows the C$^{18}$O emission with $^{13}$CO contours overlaid. Some $^{13}$CO clumps have weak or no corresponding C$^{18}$O emission. In Table\,\ref{Tab:co_clumps_1}, therefore, we only consider the sources that are detected in both the lines with main beam brightness temperatures $T_{\rm ^{13}CO}>3\sigma$ and $T_{\rm C^{18}O}>3\sigma$. The white ellipses with numbers show the extracted $^{13}$CO clumps. The average $^{13}$CO and C$^{18}$O lines within each extracted $^{13}$CO clump are presented in Fig.\,\ref{Fig:c18o-spectra}. We also display the Gaussian fitted lines with green curves. Most of the $^{13}$CO and C$^{18}$O lines can be fitted with a single-velocity Gaussian component. For the multi-velocity components, we only consider the strongest peak or the velocity components with infrared emission.

\begin{figure*}
\centering
\includegraphics[width=0.32\textwidth, angle=0]{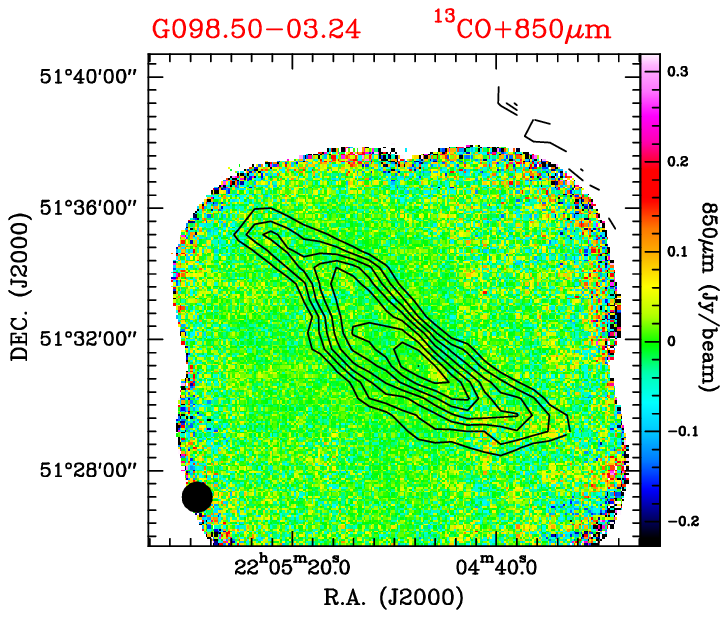}
\includegraphics[width=0.32\textwidth, angle=0]{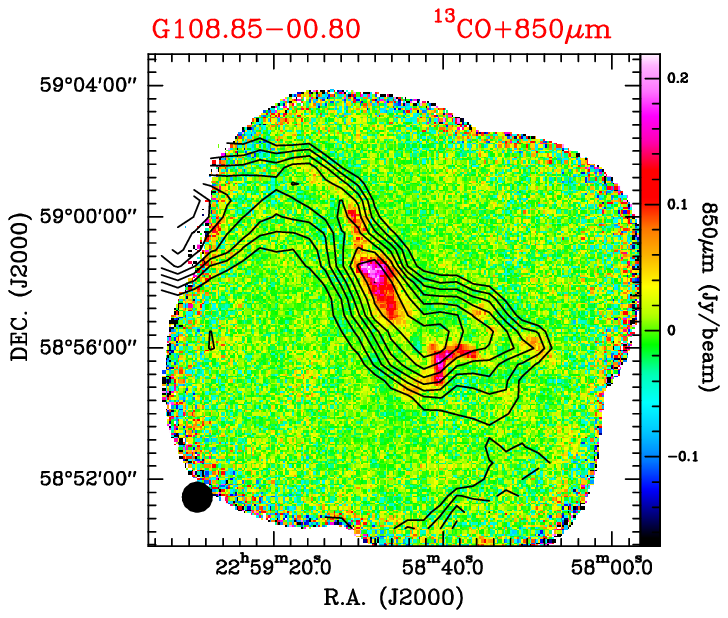}
\includegraphics[width=0.32\textwidth, angle=0]{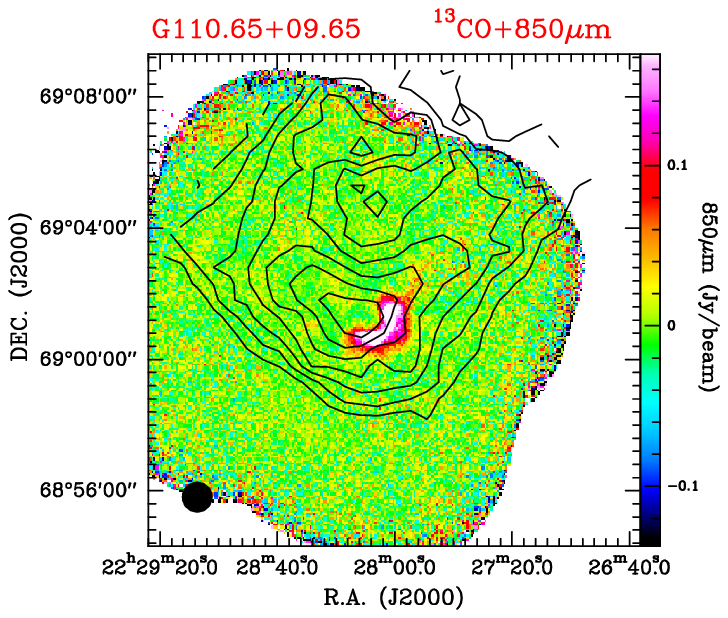}
\caption{SCUBA-2 850\,$\mu$m emission for each PGCC with overlaid $^{13}$CO contours. The contour levels of the $^{13}$CO lines are drawn at 10\% steps, starting with 30\% of the peak value. The beam size of $^{13}$CO data is indicated in the bottom-left corner. \texttt{Supplementary figures can be downloaded in https://zcp521.github.io/pub/Figs.zip.}}
\label{Fig:13co-850um}
\end{figure*}

In Fig.\,\ref{Fig:13co-c18o}, morphologically we observed that some PGCCs show clearly filamentary structure (e.g. G108.85-00.80, G116.12+08.98) and spherical structure (e.g. G115.92+09.46, G133.48+09.02), and the others are morphologically complicated and don't belong to the both cases above. The filamentary structures are ubiquitous in the Milky Way \citep{Rathborne2006,Csengeri2014,Motte2017,Zhangcp2017}. They are much elongated along the long axis of filament with aspect ratios $\gtrsim 5$ \citep{Wang2011,Wang2014}. For the spherical structure, the most massive fragments are often located at the center position of their parent clusters, with several low-mass fragments surrounding the most massive one. We find that the filamentary structures make up 23 (35.9\%) sources in the 64 PGCCs, respectively. Dense clumps elongate along their parental filament axis. The clumps in filamentary structure seem to be more compact than the others in our sample. \citet{Konyves2015} suggested that the filamentary environment is more suitable for star formation than spherical structures. \citet{Lane2016} suggested that the dense core clusters tend to be elongated, perhaps indicating a formation mechanism linked to the filamentary structure within molecular clouds.

Figure\,\ref{Fig:c18o-spectra} shows that only 56 (12.0\%) of the $^{13}$CO clumps show multi-velocity components in both $^{13}$CO and C$^{18}$O emission, and the others have single velocity components. In the direction of the second quadrant of the Milky Way, there are at most three spiral arms in line of sight, and they are located at relatively near distances without kinematic distance ambiguity. The relatively optically thick $^{13}$CO line has similar profiles to the optically thin C$^{18}$O line for most clumps, but shows a slightly broader width. Comparing the $^{12}$CO line as a dynamical tracer with the $^{13}$CO line, \citet{Zhangtw2016} also found that clumps are mostly dynamically quiescent and lack star forming activity, further indicating that the PGCCs are most likely in a very early evolutionary stage of star formation \citep{Wuyf2012,Yuan2016}.

In Figs.\,\ref{Fig:13co-12um} and \ref{Fig:13co-22um}, $^{13}$CO emission is plotted on maps of WISE 12 and 22\,$\mu$m emission. This is helpful for understanding the related infrared emission distribution in the background, and to predict the interaction relationship between the ionized gas and molecular clouds (see details in Section\,\ref{sect:ir_emission}).

\subsection{850\,$\mu$m cores}

All 64 PGCCs have been observed at 850\,$\mu$m with SCUBA-2, but only 28 (43.8\%) are detected above 5$\sigma$ ($\sigma$ is the rms noise of the image). The PGCCs G142.49+07.48 and G150.44+03.95 are not adequately covered by the 850\,$\mu$m observations, which may reduce the detection number of 850\,$\mu$m cores, but have no significantly affect on the detection statistics. The low detection rate suggests that the PGCCs have a relatively low core formation efficiency (CFE; see Section\,\ref{sect:cfe}). In total, we extracted 117 850$\mu$m cores having an effective radius range of 0.03 -- 0.48\,pc with a median value of 0.07\,pc and a detected mass range of 0.4 -- 311\,$\Msun$ with a median value of 8\,$\Msun$ for the cores. Figure\,\ref{Fig:13co-850um} shows the 850\,$\mu$m emission map (\textit{color scale}) with $^{13}$CO contours overlaid. About 26 (22.2\%) of the 117 cores have weak ($<3\sigma$) or no corresponding C$^{18}$O emission (see Fig.\,\ref{Fig:850um-spectra} and Table\,\ref{Tab:850_clumps_1}). The 850\,$\mu$m cores are strongly associated with the peak positions of C$^{18}$O emission.

\begin{figure*}
\centering
\includegraphics[width=0.32\textwidth, angle=0]{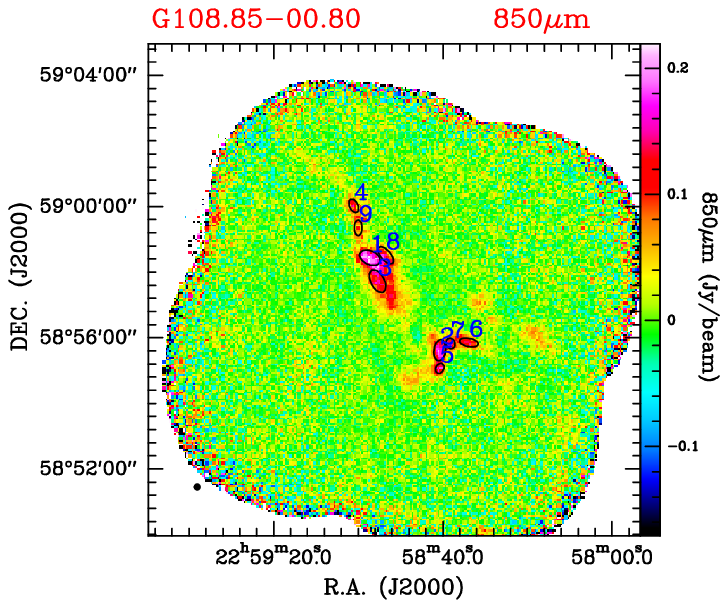}
\includegraphics[width=0.32\textwidth, angle=0]{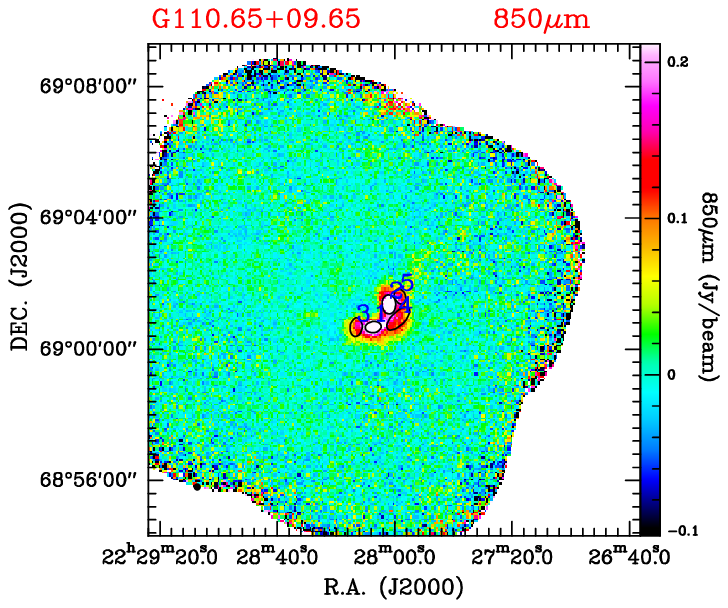}
\includegraphics[width=0.32\textwidth, angle=0]{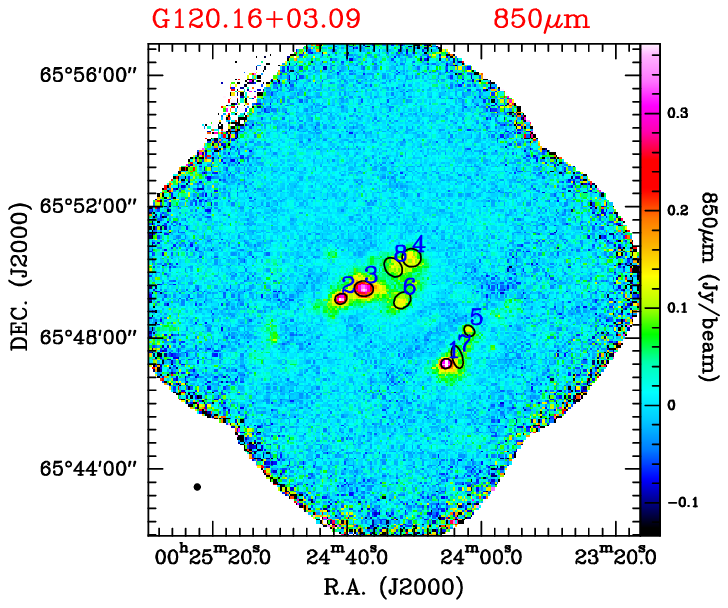}
\caption{The extracted 850\,$\mu$m cores (\textit{black ellipse}) superimposed on 850\,$\mu$m emission. The beam size of 850\,$\mu$m data is indicated in the bottom-left corner. \texttt{Supplementary figures can be downloaded in https://zcp521.github.io/pub/Figs.zip.}}
\label{Fig:850um-clumps}
\end{figure*}

In Table\,\ref{Tab:850_clumps_2}, we list the positional associations between our extracted 850\,$\mu$m cores and the AllWISE Data \citep{Cutri2013,Cutri2014}. We only search for WISE point sources within 10$''$ radii of the peak position of each 850\,$\mu$m core. In Fig.\,\ref{Fig:850um-clumps}, we show the distribution of the extracted 850\,$\mu$m cores. We find that 74 (63.2\%) of the 117 850$\mu$m cores have corresponding WISE infrared point sources, some of which may happen to be the sources in line of sight.

In Fig.\,\ref{Fig:850um-spectra}, we present the $^{13}$CO and C$^{18}$O lines extracted from the 117 850\,$\mu$m cores. We find that only 26 (22.2\%) of the 850\,$\mu$m cores show multi-velocity components in $^{13}$CO and C$^{18}$O lines, suggesting that most detections correspond to a single object along the line of sight. Compared with $^{13}$CO clumps, the 850\,$\mu$m cores have few multi-peak spectra, indicating that majority $^{13}$CO clumps at large scale are relatively more dynamically complex than the 850\,$\mu$m cores at small scale (see the error analysis in Section\,\ref{sect:error}).

\subsection{Opacity, excitation temperature, column density, and mass}

\begin{figure*}
\centering
\includegraphics[width=0.32\textwidth, angle=0]{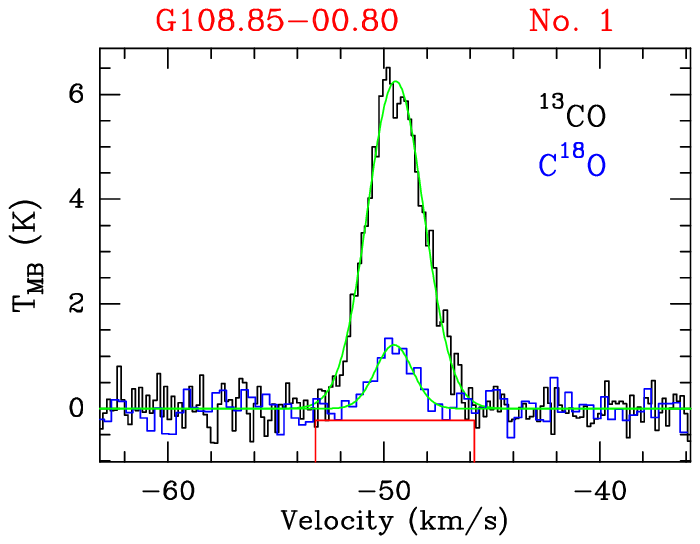}
\includegraphics[width=0.32\textwidth, angle=0]{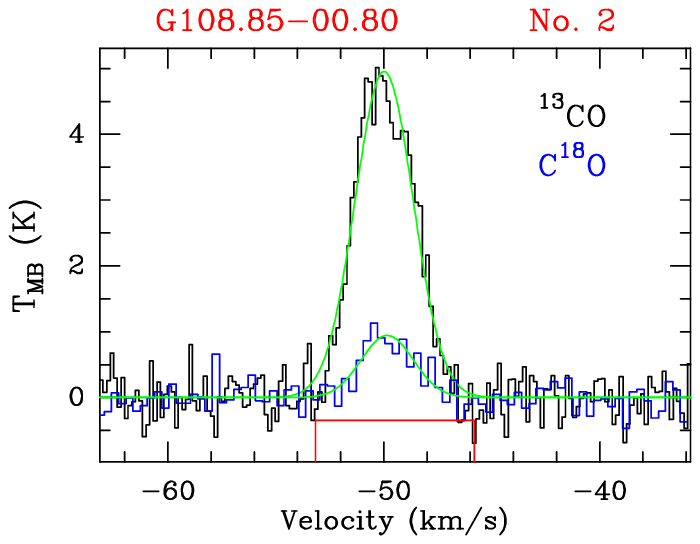}
\includegraphics[width=0.32\textwidth, angle=0]{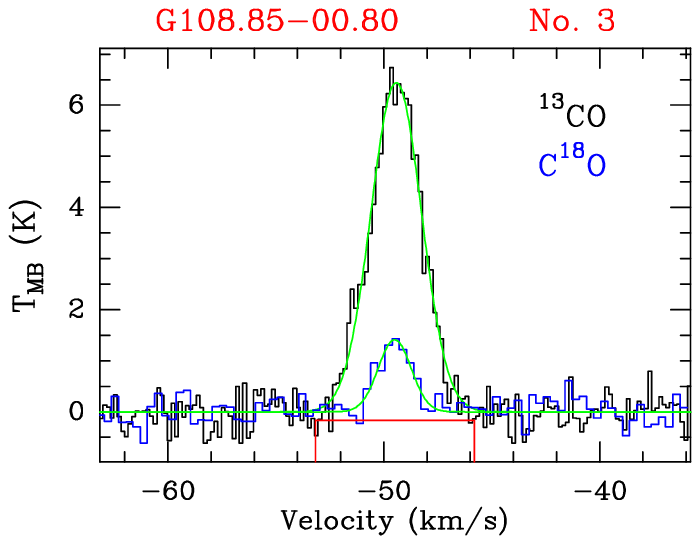}
\caption{$^{13}$CO (black line) and C$^{18}$O (blue line) lines within each extracted 850\,$\mu$m core (see Fig.\,\ref{Fig:850um-clumps}). The green lines show the Gaussian fits in each spectrum. The red window indicates the velocity range of corresponding $^{13}$CO and C$^{18}$O integrated intensity maps (see Fig.\,\ref{Fig:13co-c18o}). \texttt{Supplementary figures can be downloaded in https://zcp521.github.io/pub/Figs.zip.}}
\label{Fig:850um-spectra}
\end{figure*}

We use the approach of \citet{wong2008} to derive the opacity, excitation temperature, and column density of each clump combining $^{13}$CO and C$^{18}$O $J = 1-0$. The integrated velocity ranges for each clump are shown in Figs.\,\ref{Fig:c18o-spectra} and \ref{Fig:850um-spectra}. The relationship between opacities ($\tau$) and main-beam brightness temperatures ($T_{\rm MB}$) for $^{13}$CO and C$^{18}$O \citep{myer1983,s51} is
\begin{equation}
  \label{eq:13co2c18o}
  \frac{T_{\rm MB}(^{13}\rm CO)}{T_{\rm MB}(\rm C^{18}O)}=\frac{1-\rm exp(-\tau_{13})}{1-\rm exp(-\tau_{18})}=\frac{1 -\rm exp(-\lambda\tau_{18})}{1-\rm exp(-\tau_{18})}.
\end{equation}
Equation (\ref{eq:13co2c18o}) assumes a single excitation temperature for both molecules and throughout the line of sight, and assumes $\tau_{13} = \lambda \tau_{18}$, where $\lambda$ is the abundance ratio between $\rm ^{13}C^{16}O$ and $\rm ^{12}C^{18}O$. $\lambda$ can be derived from relation in \citet{wils1994} and \citet{Pineda2013} to be as
\begin{equation}
  \lambda = \frac{[\rm ^{13}C^{16}O]}{[\rm ^{12}C^{18}O]}=\frac{58.8R_{\rm GC}+37.1}{4.7R_{\rm GC}+25.05},
\end{equation}
where $R_{\rm GC}$ is the Galactocentric distance. We only consider the sources detected with main beam brightness temperatures $T_{\rm ^{13}CO}>3\sigma$ and $T_{\rm C^{18}O}>3\sigma$ (see Table\,\ref{Tab:co_clumps_2}). Furthermore, the excitation temperature $T_{\rm ex}$ is derived from the radiative transfer equation:
\begin{equation}
     J(T)= {\rm T_{0}}/[{\rm exp(T}_{0}/T)-1]
\end{equation}
\begin{equation}
     T_{\rm MB}=f[J(T_{\rm ex})-J(T_{\rm bg})][1-\rm exp(-\tau)]
\end{equation}
where $f$ is the beam filling factor which we assume as $f=1$, $T_{bg}$ = 2.73\,K is the cosmic microwave background temperature and $T_{0}=h\nu/k$ = 5.29\,K for the $J=1-0$ transition of $^{13}$CO \citep{wong2008}. We then obtain the molecular $^{13}$CO column density $N(^{13}$CO) from the relation \citep{bour1997}:
	\begin{eqnarray}
   \label{eq:tmb}
    N(^{13}{\rm CO})_{\rm thin}= \frac{T_{\rm ex}+0.88}{1-{\rm exp}(-5.29/T_{\rm ex})} \nonumber \\
    \times \frac{2.42\times10^{14}}{J(T_{\rm ex})-J(T_{\rm bg})}\int T_{\rm MB}(^{13}{\rm CO})dv,
	\end{eqnarray}
where $N(^{13}{\rm CO})_{\rm thin}$ and $v$ are in units of cm$^{-2}$ and $\kms$, respectively. We then apply a correction factor $\rm \tau/(1-exp(-\tau))$ to the $^{13}$CO column density \citep{Pineda2010,Liu2013}:
\begin{equation}
    \label{eq:tmb}
    N'(^{13}{\rm CO})_{\rm corrected}=N(^{13}{\rm CO})_{\rm thin}\times \rm \frac{\tau_{13}}{1-exp(-\tau_{13})}
\end{equation}
Finally, the molecular hydrogen column $N_{\rm H_2}$ was calculated, assuming that the [H$_{2}$/$^{13}$CO] abundance ratio is $7\times10^{5}$ \citep{frer1982}.

The $^{13}$CO clump mass is given by the integral of the column density across the source via formula \citep{Kauffmann2008}:
	\begin{eqnarray}
	\label{equa_kappa}
M_{\rm H_2}^{^{13}{\rm CO}}=\mu_{\rm H_2}m_{\rm H}D^2\int N_{\rm H_2} d\Omega,
	\end{eqnarray}
where $\mu_\mathrm{H_2}=2.8$, $m_\mathrm{H}=1.008$\,u, $D$, and $\Omega$ are the mean molecular weight, the mass of a hydrogen atom, the distance, and the solid angle of the source, respectively. The masses of all extracted $^{13}$CO clumps are listed in Table\,\ref{Tab:co_clumps_2}.

\subsection{Virial analysis}
\label{sect:virial}

The virial theorem can be used to test whether fragments are in a stable state. Under the assumption of a simple spherical fragment with a density distribution of $\rho = r^{-2} $, ignoring magnetic fields and bulk motions of the gas, the virial mass of a fragment can be estimated from the formula \citep{MacLaren1988,Evans1999}:
	\begin{eqnarray}
	\label{equa_virial-mass}
    M_{\rm vir} \simeq 126\, R_{\rm eff}\, \Delta V^{2}_{\rm C^{18}O}\, (\Msun),
	\end{eqnarray}
where $R_{\rm eff} = \rm FWHM/(2\sqrt{ln2})$ is the effective radius of the fragment in pc, and $\Delta V_{\rm C^{18}O}$ (listed in Tables \ref{Tab:co_clumps_1} and \ref{Tab:850_clumps_1}) is the FWHM of the line profile in $\kms$. $\Delta V_{\rm C^{18}O}$ is the measured C$^{18}$O linewidth using Gaussian line fitting. For a typical cold cloud ($\rm <20\,K$), the thermal width is only a few tenths narrower than the observed linewidth, thus the observed linewidth is presumed to be representative of the turbulent velocity structure. The spatial resolution of the C$^{18}$O data is somewhat larger than the sizes of individual cores, hence we just consider the C$^{18}$O spectrum within one pixel corresponding to the peak position of each core (see error analysis in Section\,\ref{sect:error}). The virial parameter $\alpha_{\rm vir}$ is defined by $\alpha_{\rm vir} = M_{\rm vir}/M$. The virial masses and virial parameters are listed in Tables \ref{Tab:co_clumps_2} and \ref{Tab:850_clumps_2}.

\subsection{Dust mass and surface density}
\label{sect_mass}

We assume that the dust emission is optically thin and the gas-to-dust ratio is 100. The fragment masses are calculated using dust opacity $\kappa_{\nu}$ = 0.0182 cm$^2$\,g$^{-1}$ at 850\,$\mu$m \citep{Kauffmann2008} assuming a gas to dust mass ratio of 100 for a model of dust grains with thin ice mantles at a gas density of 10$^6$ cm$^{-3}$\citep{Ossenkopf1994}. The total mass, $M_{\rm H_2}^{\rm 850{\mu}m}$, of the 850\,$\mu$m sources can therefore be calculated via the formula \citep{Kauffmann2008}:
	\begin{eqnarray}
	\label{equa_mass}
	\left(\frac{M_{\rm H_2}^{\rm 850{\mu}m}}{\Msun}\right) = 0.12 \left({\rm e}^{14.39\left(\frac{\lambda}{\rm mm}\right)^{-1}\left(\frac{T_{\rm dust}}{\rm K}\right)^{-1}}-1\right) \nonumber \\
	\times \left(\frac{\kappa_{\nu}}{\rm cm^{2}g^{-1}}\right)^{-1} \left(\frac{S_{\nu}}{\rm Jy}\right) \left(\frac{D}{\rm kpc}\right)^{2} \left(\frac{\lambda}{\rm mm}\right)^{3},
	\end{eqnarray}
where $\lambda$ is the observed wavelength in mm, $T_{\rm dust}$ is the dust temperature in K, $S_{\nu}$ is the integrated flux in Jy, and $D$ is the distance to the source in kpc. For all fragments, we adopt the associated excitation temperature of $^{13}$CO $J=1-0$ as an approximated dust temperature \citep{Liu2013}. The surface density ($\Sigma$) can be derived from $\Sigma = M/(\pi R^{2}_{\rm eff})$ in units of $\rm g\,cm^{-2}$, and here $R_{\rm eff}$ is the effective radius of the fragment, and FWHM is the source size. These corresponding parameters are also listed in Tables \ref{Tab:850_clumps_1} and \ref{Tab:850_clumps_2}.

\section{Discussion}
\label{sect:discussion}

We have surveyed 64 PGCCs with CO and in the 850\,$\mu$m continuum in the second quadrant of the Milky Way. The CO observations have low spatial resolution, and therefore trace relatively extended molecular clouds and clumps at larger scales, whilst the 850\,$\mu$m-continuum observations have relatively high spatial resolution and are used to trace the dense cores at smaller scales, embedded within the molecular clouds. Investigating the fragments at different scales, and comparing their differences will help us to improve our understanding of the early stages of the star-formation process. By combining CO isotopologues ($^{13}$CO and C$^{18}$O), important physical parameters can be quantitatively estimated to characterize and increase our knowledge of the clump properties. 5 of the PGCCs, G108.85-00.80, G112.52+08.38, G120.67+02.66, G120.98+02.66, and G121.92-01.71, are distributed over larger regions than the scan map size ($14'\times14'$) and further observations over a larger region are needed for a complete analysis.

\subsection{Error analysis of different beam sizes}
\label{sect:error}

The CO and 850\,$\mu$m observations have beam sizes of around 52$''$ and 14$''$, respectively. To derive some parameters of 850\,$\mu$m cores, such as velocity dispersion and temperature, we have to use the CO line data observations for estimation. However, there is no simple way of assigning CO emission to each 850\,$\mu$m core when they form a tight cluster, for example, in G172.85+02.27. That is, the CO emission toward each one of these six 850\,$\mu$m cores is contaminated by emission from nearby cores. If the 850\,$\mu$m cores are not located in a tight cluster, it seems that we can use a reasonable filling factor, $f=(14/52)^2$, to estimate the excitation temperature assuming all integrated-intensity of $^{13}$CO clump emits from the dense and isolated 850\,$\mu$m core. However, the large 52$''$ beam means that the velocity gradients from, e.g., accretion along filaments, rotation, and even molecular outflows, will overestimate the velocity dispersion of each 850\,$\mu$m core. Therefore, this will lead to high uncertainties for our estimation. The velocity dispersion, virial mass and virial parameter of the 850\,$\mu$m cores will be overestimated, and the excitation temperature at the position of each 850\,$\mu$m core will be underestimated.

\subsection{Infrared emission}
\label{sect:ir_emission}

The extended 12\,$\mu$m emission originates mainly from polycyclic aromatic hydrocarbons \citep{wats2008}, which are excited by UV radiation at the interface between the expanding \HII region and the ambient interstellar medium \citep{Zhang2016}. The extended 22\,$\mu$m emission is mostly produced by relatively hot dust \citep{ander2012,Faimali2012}, and is a good tracer of early star formation activity.

Figures\,\ref{Fig:13co-12um} and \ref{Fig:13co-22um} compare infrared emission of WISE 12 and 22\,$\mu$m with the $^{13}$CO emission contours. The morphological distribution of the $^{13}$CO emission is correlated or uncorrelated with the 12 and 22\,$\mu$m emission, which correspond to infrared-bright and infrared-dark PGCCs, respectively. We find that $\sim$\,30\% of PGCCs are infrared bright after visually inspecting the image, whilst $\sim$\,70\% sources are infrared dark. We also note that $\sim$\,15\% of the infrared dark PGCCs have more than one infrared bright core, which are also correlated with a peak in the $^{13}$CO emission. Positionally matching the AllWISE Catalog \citep{Cutri2013,Cutri2014} with the extracted 850\,$\mu$m sources (see Table\,\ref{Tab:850_clumps_2}), we find that 74 of the 117 850$\mu$m cores have corresponding WISE infrared point sources. Those with or without infrared point sources may be protostellar or infrared quiet/starless cores, respectively \citep{Yuan2017}. This suggests that the infrared dark PGCCs with infrared bright cores are more evolved, but are younger than the infrared bright PGCCs.

We find no infrared dust bubbles \citep{chur2006,chur2007}, which would show strong 22\,$\mu$m emission surrounded by a ringlike 12\,$\mu$m emission shell. The 12 and 22\,$\mu$m emission have no significant morphological differences. Compact \HII regions or bright infrared cores may have an effect on the evolution of early star formation \citep{Zhang2017}. The integrated-intensity maps of some PGCCs show steep gradients, e.g. G098.50-03.24, G127.88+02.66, G151.08+04.46 (see Fig.\,\ref{Fig:13co-c18o}). This suggests that the molecular clouds have been compressed by nearby warm clouds, such as G035.39-00.33 \citep{Liu2018}, possibly indicating cloud-cloud collisions, such as in the case of G178.28-00.61 (Zhang et al., in prep.).

\subsection{Fragmentation}

In Fig.\,\ref{Fig:c18o-spectra}, we present the $^{13}$CO and C$^{18}$O line spectra extracted from the identified 468 $^{13}$CO clumps in our sample of 64 PGCCs. They typically have sizes of $\sim$\,0.2 -- 2\,pc. This shows that each of the PGCCs fragments into an average of $\sim$\,7.3 $^{13}$CO clumps. The G144.84+00.76 PGCC fragments into 14 clumps. We suggest that the fragmentation is ubiquitous and a necessary process of the early stage of star formation. Analysis shows that most of the clumps are associated with CO structures. Only 68 sources of the $^{13}$CO clumps were not detected with C$^{18}$O lines, suggesting that most of the clumps are relatively dense.

Figure\,\ref{Fig:850um-clumps} shows examples of 850\,$\mu$m emission maps with extracted cores overlaid. In total, 117 cores are extracted at 850\,$\mu$m. Less than half (28) of the 64 PGCCs have been detected with 850\,$\mu$m continuum, indicating that each PGCC fragments into 4.2 cores on average with an effective radius of 0.03 -- 0.48\,pc. We suggest that the number of the fragments is strongly associated with the fragment size, and the results might also be dependent on the sensitivity \citep{Pokhrel2017}.

\subsection{Mass-size relation}

\begin{figure}[h]
\centering
\includegraphics[width=0.45\textwidth, angle=0]{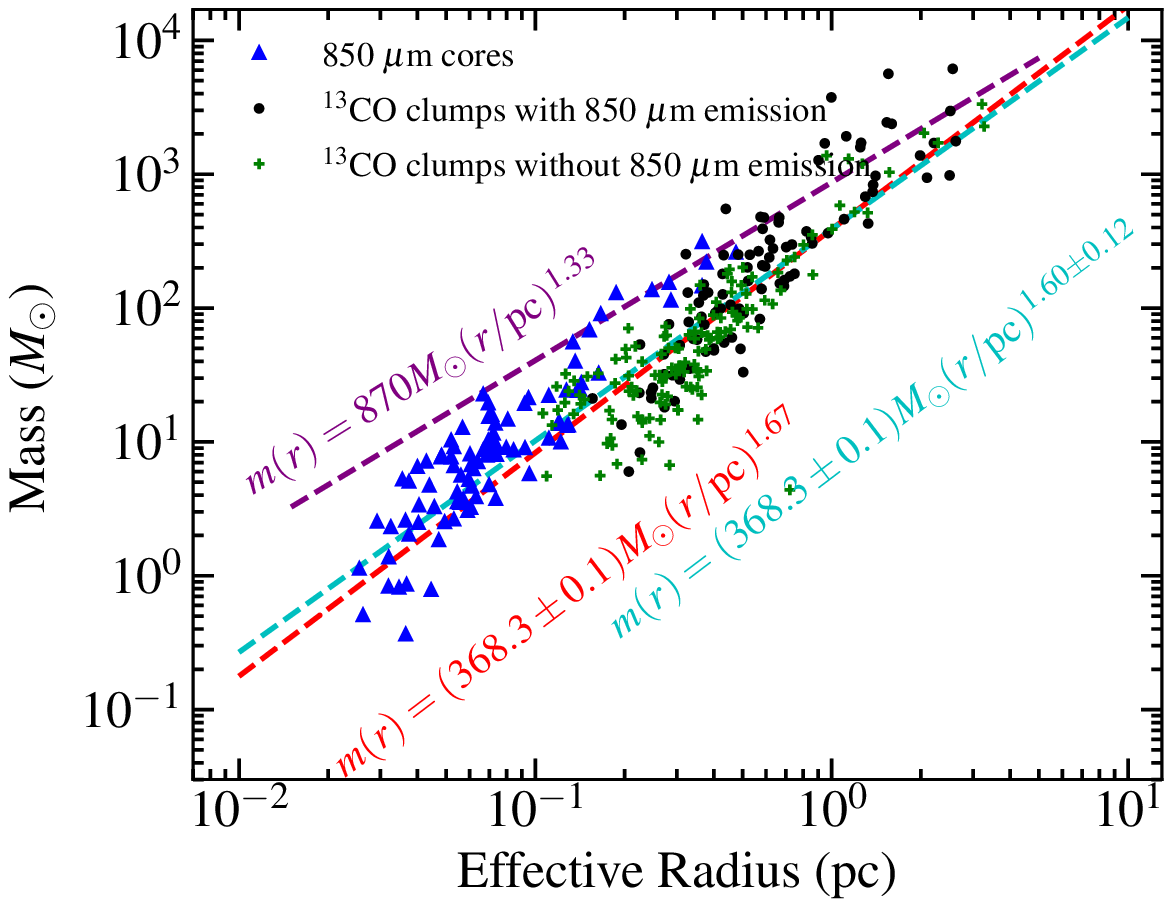}
\caption{\textit{Mass-Radius} distributions of Gaussian 850\,$\mu$m cores (\textit{blue triangles}), $^{13}$CO clumps with (\textit{black dots}) and without (\textit{green crosses}) 850\,$\mu$m emission. Their masses and effective radii are listed in Tables \ref{Tab:co_clumps_2} and \ref{Tab:850_clumps_2}, respectively. The purple line delineates the threshold introduced by \citet{Kauffmann2010}, separating the regimes into high-mass and low-mass star forming candidates. The red line shows a power-law fit using least-squares fitting in log-space with a fixed exponent 1.67 to the mass-size relation for clumps that undergo quasi-isolated gravitational collapse in a turbulent medium \citep{Ligx2017,ZhangLi2017}. The cyan line presents a power-law fit of all the data point using least-squares fitting. The corresponding formulas are also shown nearby the lines. }
\label{Fig:mass_size}
\end{figure}

\citet{Ligx2017} derived a scaling relation of $m \sim r^{5/3}$ to describe the properties of the gravitationally bound structures, where the multiplication factor of the relationship is determined by the level of ambient turbulence. A higher level of turbulence leads to a higher mass at a given scale. It has been found that the scaling provides a good description to the fragments observed from sub-pc scales to those of a few pc \citep{Zhangcp2017}. In Fig.\,\ref{Fig:mass_size}, the red dashed line shows the mass-size relation derived assuming $m\sim r^{5/3}$ from \citet{Ligx2017} and \citet{Zhangcp2017}. Figure\,\ref{Fig:mass_size} also displays the mass-size relation derived from a linear fit to the data. In general, the results obtained from these fits are very similar. This clearly demonstrates that these clumps do not obey ``Larson's third law'', where the power-law exponent is $\sim$ 1.9 \citep{Larson1981,Solomon1987}, but is consistent with the prediction made in \citet{Ligx2017}.

Having established the relation, it is possible to use the properties of the observed clumps to estimate the turbulence in the ambient medium. In our sample, we found that our structures satisfy the relationship, $m(r) = (368.3\pm0.1) M_{\odot} (r/ {\rm pc})^{1.67}$ (see Figure\,\ref{Fig:mass_size}). The multiplication factor is small compared to that found in high-mass star-forming regions by \citet{Urquhart2014}, who had a relationship of $m(r) = 2630 M_{\odot} (r/ {\rm pc})^{1.67}$, and \citet{Zhangcp2017}, $m(r) = 7079 M_{\odot} (r/ {\rm pc})^{1.67}$. The sample in \citet{Zhangcp2017} are more massive and denser than those in \citet{Urquhart2014}. At a given scale, the masses of gas condensation in our PGCC sample are around 1/10 of that of typical Galactic high-mass star-forming regions \citep{Urquhart2014}. Using the scaling relation presented in \citet{Ligx2017}, this implies that the energy dissipation rate of the ambient turbulence should be $1/30$ of that of the Galactic massive star-forming regions, where we expect the observed velocity dispersion of the molecular gas in our PGCC sample to be 1/3 times of the averaged Galactic value on a given scale\footnote{These numbers are obtained using the scaling relations presented in  \citet{Ligx2017}, where the mass-size relation is determined by $m \approx \epsilon_{\rm cascade}^{2/3} \eta^{-2/3} G^{-1} r^{5/3}$ (where $m$ is the critical mass, $r$ is the source size, and $\epsilon_{\rm cascade}\approx \eta \sigma_{\rm v}^3 /l $ is the turbulence energy dissipation rate of the ambient medium).}. In general, the level of turbulence in PGCC sample is significantly lower than the Galactic average. This is consistent with our previous findings in \citet{Zhangtw2016}.

\subsection{Low-mass star formation}

\begin{figure}[h]
\centering
\includegraphics[width=0.45\textwidth, angle=0]{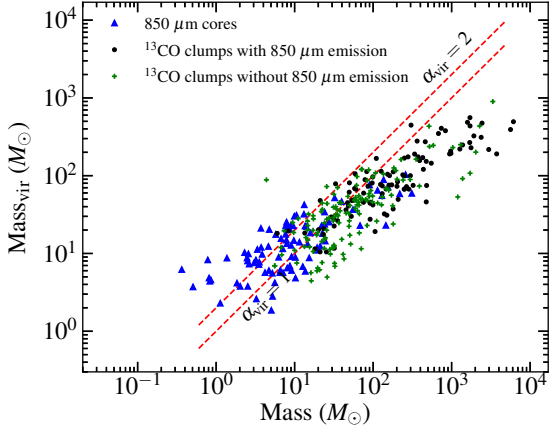}
\caption{\textit{Mass$_{\rm vir}$-Mass} distributions of Gaussian 850\,$\mu$m cores (\textit{blue triangles}), $^{13}$CO clumps with (\textit{black dots}) and without (\textit{green crosses}) 850\,$\mu$m emission. The parameters are listed in Tables \ref{Tab:co_clumps_2} and \ref{Tab:850_clumps_2}. Two dashed lines delineate the thresholds of $\rm \alpha_{vir}=1$ and $\alpha_{\rm vir}=2$. }
\label{Fig:mass_mass}
\end{figure}

In Fig.\,\ref{Fig:mass_size}, we present the mass-size plane for the extracted $^{13}$CO clumps and 850\,$\mu$m cores. Comparison with the high-mass star formation threshold of $m(r) > 870 {\Msun} (r/{\rm pc})^{1.33}$ empirically proposed by \citet{Kauffmann2010} allows us to determine whether these fragments are capable of giving birth to massive stars. The data points are mostly distributed below the threshold, given by the purple dashed line. Therefore, it appears that the majority of $^{13}$CO clumps and 850\,$\mu$m cores are low-mass star-forming region candidates.

In Fig.\,\ref{Fig:mass_mass}, we present virial mass vs. fragment mass distributions for the $^{13}$CO clumps and 850\,$\mu$m cores. Two dashed lines show the thresholds with virial parameters $\alpha_{\rm vir} = 1$ and $\alpha_{\rm vir} = 2$. We find that $\sim$\,26\% of $^{13}$CO clumps have $\alpha_{\rm vir} > 1$, and $\sim$\,5\% have $\alpha_{\rm vir} > 2$, whilst $\sim$\,71\% of the 850\,$\mu$m cores have $\alpha_{\rm vir} > 1$, with $\sim$\,37\% having $\alpha_{\rm vir} > 2$. This indicates that most of the 850\,$\mu$m cores are gravitationally unbound and are either stable or expanding \citep{Hindson2013}, relatively to the $^{13}$CO clumps. It is also likely that kinetic energy is larger than the gravitational energy, suggesting that such cores have to be confined by some external pressure \citep{Bertoldi1992,Pattle2015}. Therefore, a long timescale for star formation is required for most of our local PGCCs, or they will never form stars.

Mass surface density, $\Sigma$, is a commonly used parameter to assess the high-mass star formation potential. \citet{Urquhart2014} suggested that the surface density of 0.05 $\rm g\,cm^{-2}$ might represent a minimum threshold of efficient massive star formation, as is suitable for pc-scale clumps. According to this threshold, parts of the $^{13}$CO clumps are potential candidates of massive star formation. However, we note that most of the candidates have a typical size less than 1.0\,pc. \citet{Traficante2017} argued that $\rm \Sigma = 0.12\,g\,cm^{-2}$ may represent the minimum surface density at clump scales for high-mass star formation to occur, based on the analysis of dynamic activity associated with their parent clump. \citet{Krumholz2008} suggest that a minimum mass surface density of 1\,g\,cm$^{-2}$ is required to prevent fragmentation into low-mass cores through radiative feedback, thus allowing high-mass star formation. For the $^{13}$CO clumps and 850\,$\mu$m cores in this work, we find that the mean values of surface densities are 0.13 and 0.39\,g\,cm$^{-2}$, respectively (see Figure\,\ref{Fig:relations}). Therefore, the surface densities further prove that some of $^{13}$CO clumps and 850\,$\mu$m cores have potential to form high-mass stars but the majority would form low-mass stars.

\subsection{Core mass function}

\begin{figure}[h]
\centering
\includegraphics[width=0.45\textwidth, angle=0]{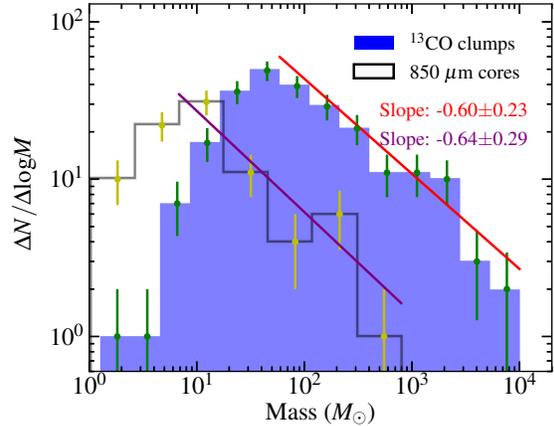}
\caption{Mass distribution for $^{13}$CO clumps (\textit{blue}) and 850\,$\mu$m cores (\textit{black}). The parameters are listed in Tables\,\ref{Tab:co_clumps_2} and \ref{Tab:850_clumps_2}. The slopes of the fitted power law index are shown in the histogram. The error bars represent the standard deviation of a Poisson distribution $\sqrt{\Delta N/\Delta{\rm log} M}$. }
\label{Fig:mass-spectum}
\end{figure}

The core mass function (CMF) generally has a comparable slope with the stellar initial mass function and, consequently, the Salpeter power-law with a logarithmic slope of -1.35 \citep{Zinnecker2007,Salpeter1955}. Previous observations show that massive stars usually form in dense clusters, so competitive accretion of protostars from their common gas reservoir was used to explain the observed Salpeter stellar mass distribution for massive stars \citep{Bonnell2001,Klessen2001}. To investigate the intermediate- and high-mass star formation, in Fig.\,\ref{Fig:mass-spectum}, we simply fit the mass spectra in the mass ranges between 400 and 8000\,$\Msun$ for $^{13}$CO clumps and between 7 and 800\,$\Msun$ for 850\,$\mu$m cores with a linear least square method using the obtained clump and core masses in our observations, respectively. The lower mass limit used to define the power law tail is derived from the peak positions (at $\sim$400\,$\Msun$ for the clumps, and $\sim$7\,$\Msun$ for the cores) from low-mass to high-mass end of the mass spectra distributions in Figure\,\ref{Fig:mass-spectum}. The derived two slopes of the mass spectrum are similar to each other with clump scope $k_{\rm clump} = -0.60\pm0.23$ and core slope $k_{\rm core} = -0.64\pm0.29$, which are much flatter than the Salpeter stellar initial mass function and the CMFs of massive star-forming candidates \citep[e.g.,][]{Beuther2004,Bontemps2010,Ohashi2016,Csengeri2017}. For low-mass star-forming objects, \citet{Marsh2016} obtained a slope of $-0.55\pm0.07$ in the Taurus L1495 cloud, \citet{Elia2013} derived a slope of $-0.7 \pm 0.3$ for the gas clump distribution in the third Galaxy quadrant, and \citet{Kim2004} also derived a shallower mass function slope of $-0.59 \pm 0.32$ for their clump sample named CMa\,OB1 and G220.8-1.7. The three cases above are consistent with our results. The similar slopes may be resulted from their similar initial conditions. We also have to note that the sample distribution at different distances and the contamination from large scale structure may lead to uncertain slopes \citep{Moore2007,Reid2010}.

\subsection{Core formation efficiency}
\label{sect:cfe}

The core formation efficiency (CFE) describes the fraction of clump mass that has converted into denser cores \citep{Elia2013,Veneziani2017}. Hence the CFE is defined as:
	\begin{eqnarray}
	\label{equa_cfe}
   {\rm CFE} = \frac{M_{\rm core}}{M_{\rm core}+M_{\rm clump}},
	\end{eqnarray}
where $M_{\rm core}$ is the mass of 850\,$\mu$m cores, and $M_{\rm clump}$ is mass of the $^{13}$CO clump that hosts those associated 850\,$\mu$m cores. Considering that $M_{\rm clump}$ is estimated from the extracted Gaussian clumps, the diffuse gas component of the cloud will be missing. Additionally, the clump masses are estimated by $^{13}$CO which will be depleted in low temperatures ($\rm < 18\,K$) \citep{Pillai2007,Pillai2011}, hence, $M_{\rm clump}$ will be underestimated. The cores in our sample are considered to be gravitationally bound objects. Using the core and clump masses of the entire sample to estimate the CFE, we get a CFE of 3.0\%\footnote{Here we consider all the extracted clumps and cores.}. Of all 64 PGCCs, only 28 (43.8\%) are detected at 850\,$\mu$m with emission above 5$\sigma$, indicating a low CFE. Our estimated CFE is much lower than those estimated from the conversion of molecular clouds to clumps across the first quadrant (5 -- 8\%; \citealt{Eden2012,Eden2013}); the first and second quadrants (5 -- 23\%; \citealt{Battisti2014}); the fourth quadrant (8 -- 39\%; \citealt{Veneziani2017}), and the Galactic Centre (10 -- 13\%; \citealt{Csengeri2016a}).

\begin{figure*}[h]
\centering
\includegraphics[width=0.32\textwidth, angle=0]{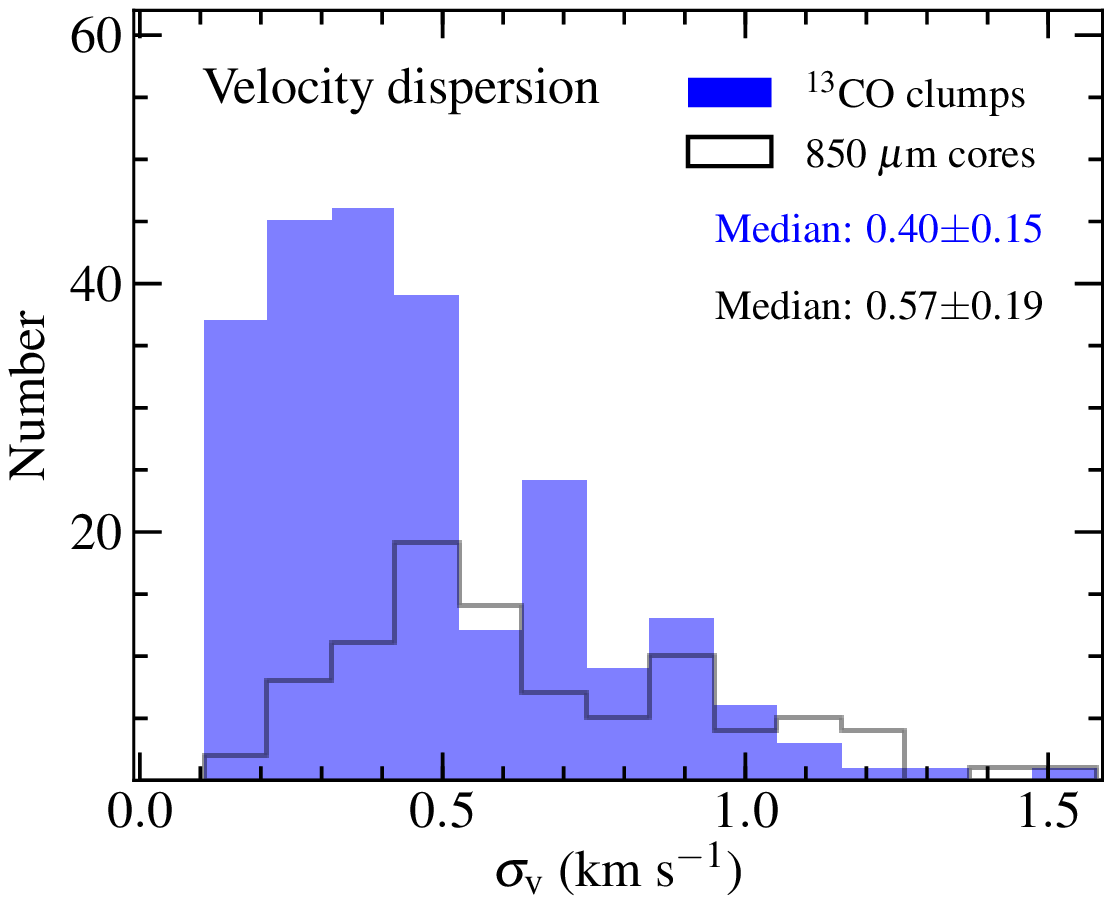}
\includegraphics[width=0.32\textwidth, angle=0]{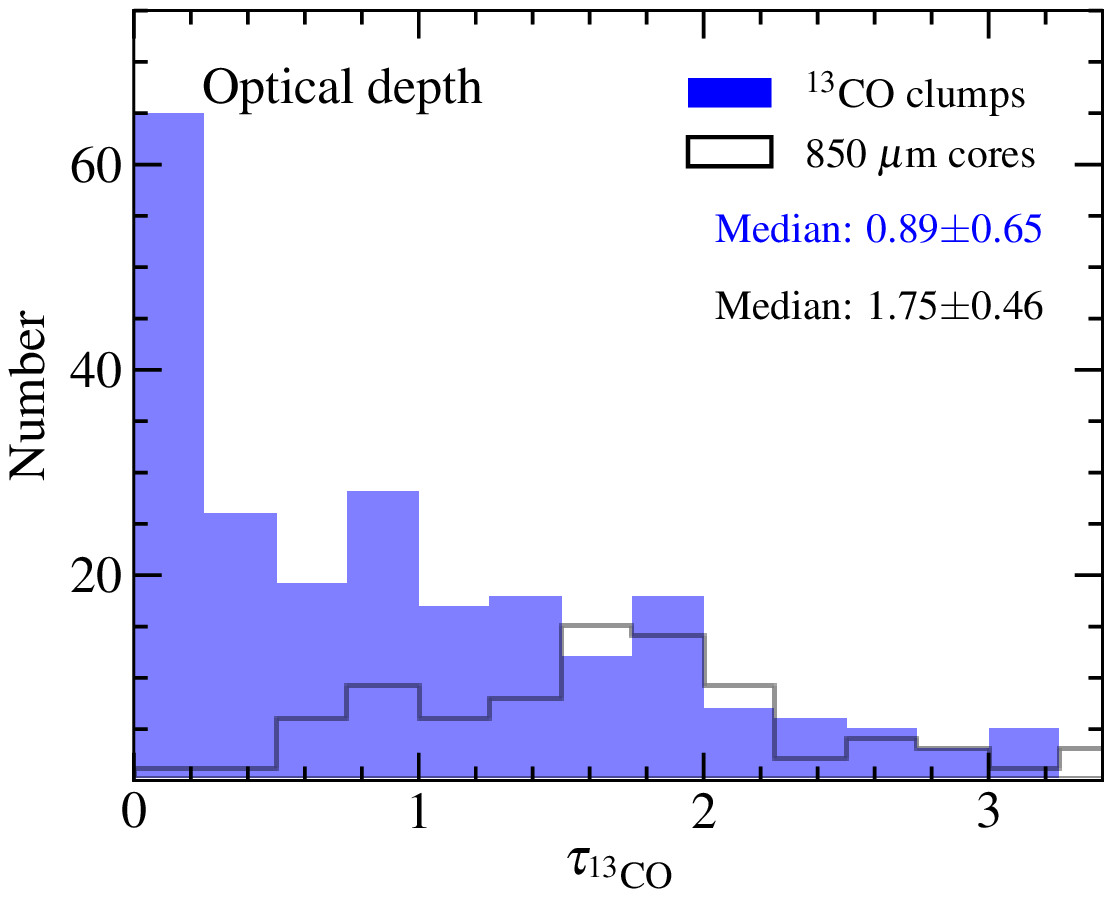}
\includegraphics[width=0.32\textwidth, angle=0]{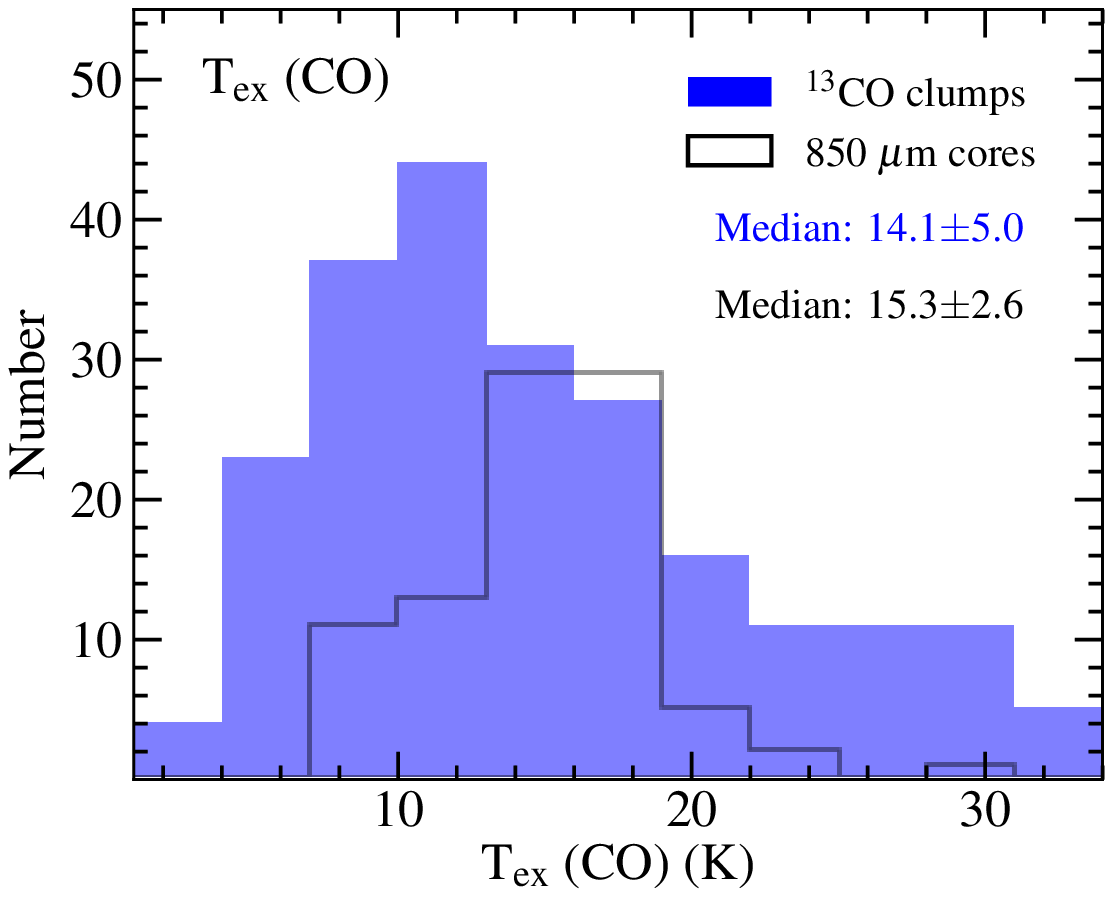}
\includegraphics[width=0.32\textwidth, angle=0]{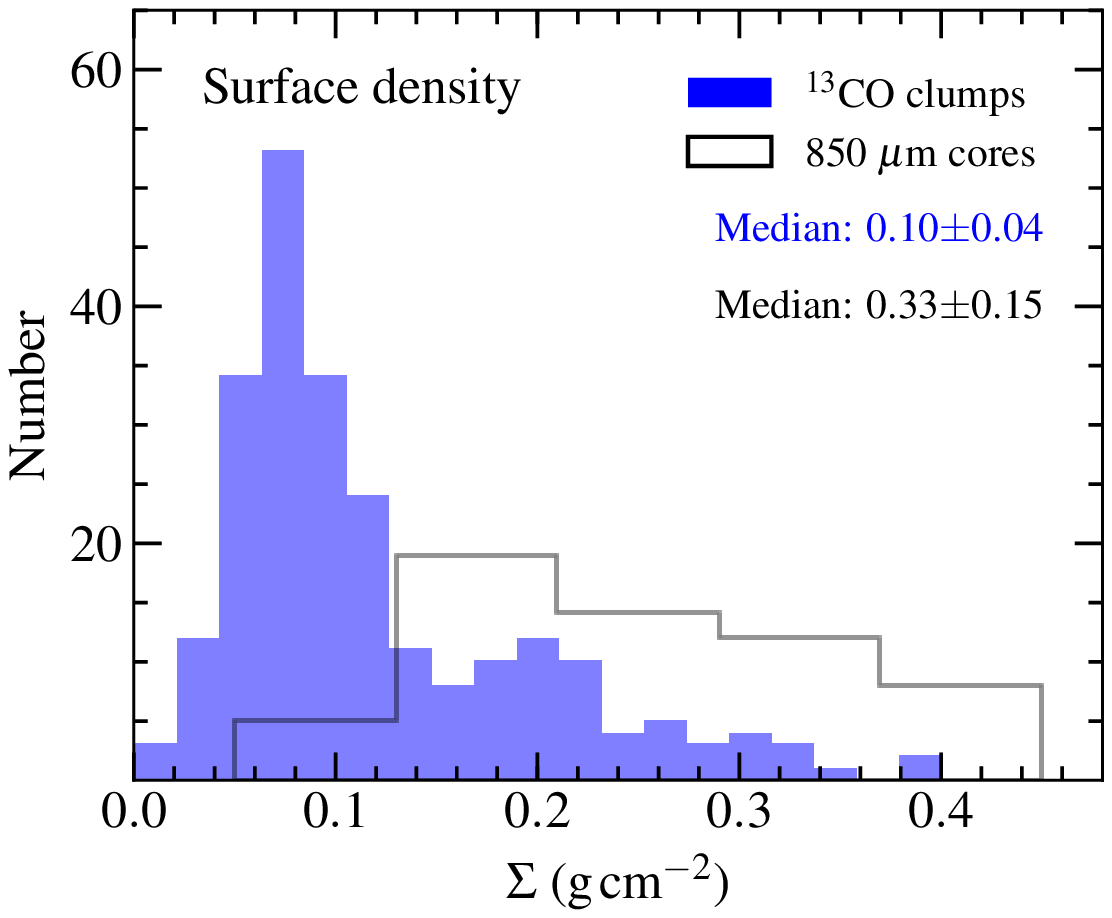}
\includegraphics[width=0.32\textwidth, angle=0]{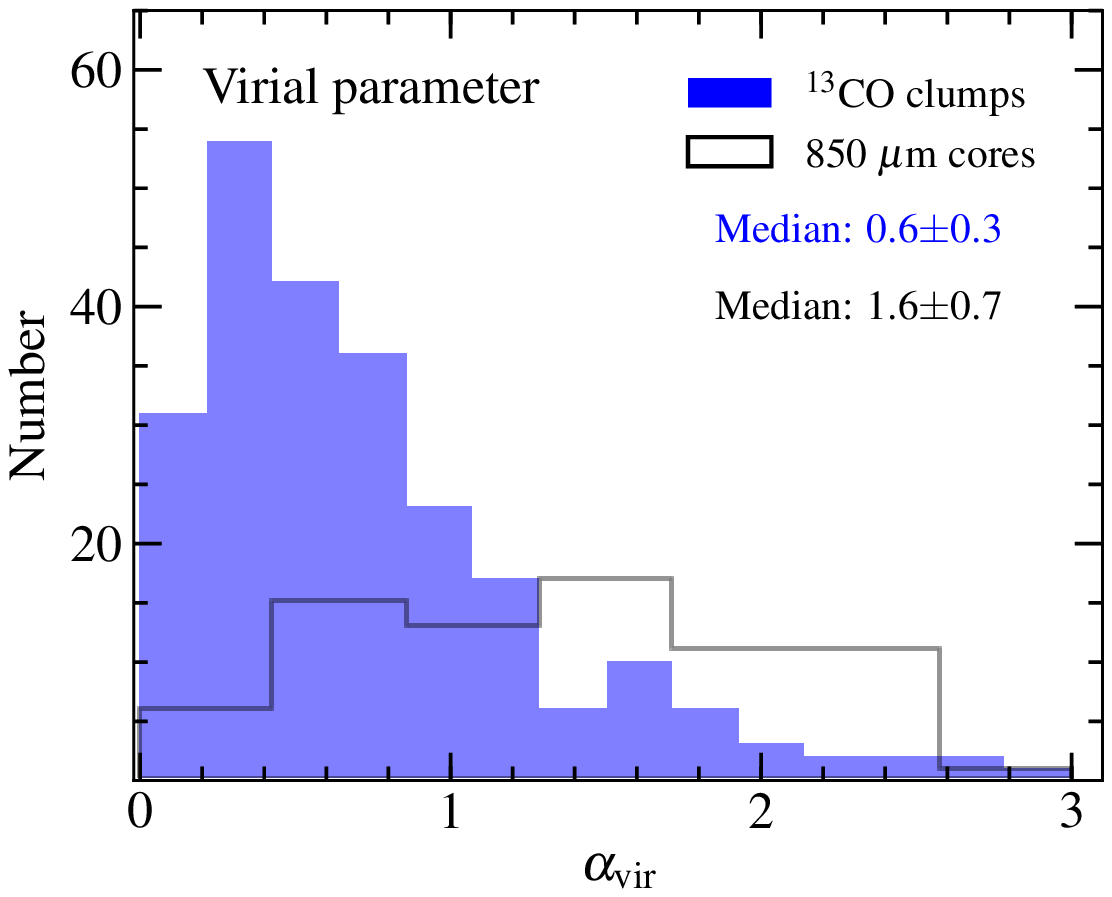}
\caption{Histograms of velocity dispersion, optical depth, excitation temperature, surface density, and virial parameter for $^{13}$CO clumps (\textit{blue}) and 850\,$\mu$m cores (\textit{black}). The parameters are listed in Tables \ref{Tab:co_clumps_1}, \ref{Tab:co_clumps_2}, \ref{Tab:850_clumps_1}, and \ref{Tab:850_clumps_2}. The corresponding median value is presented in each frame. The uncertainty on each median calculated represents median absolute deviation.}
\label{Fig:relations}
\end{figure*}

\begin{figure*}[h]
\centering
\includegraphics[width=0.32\textwidth, angle=0]{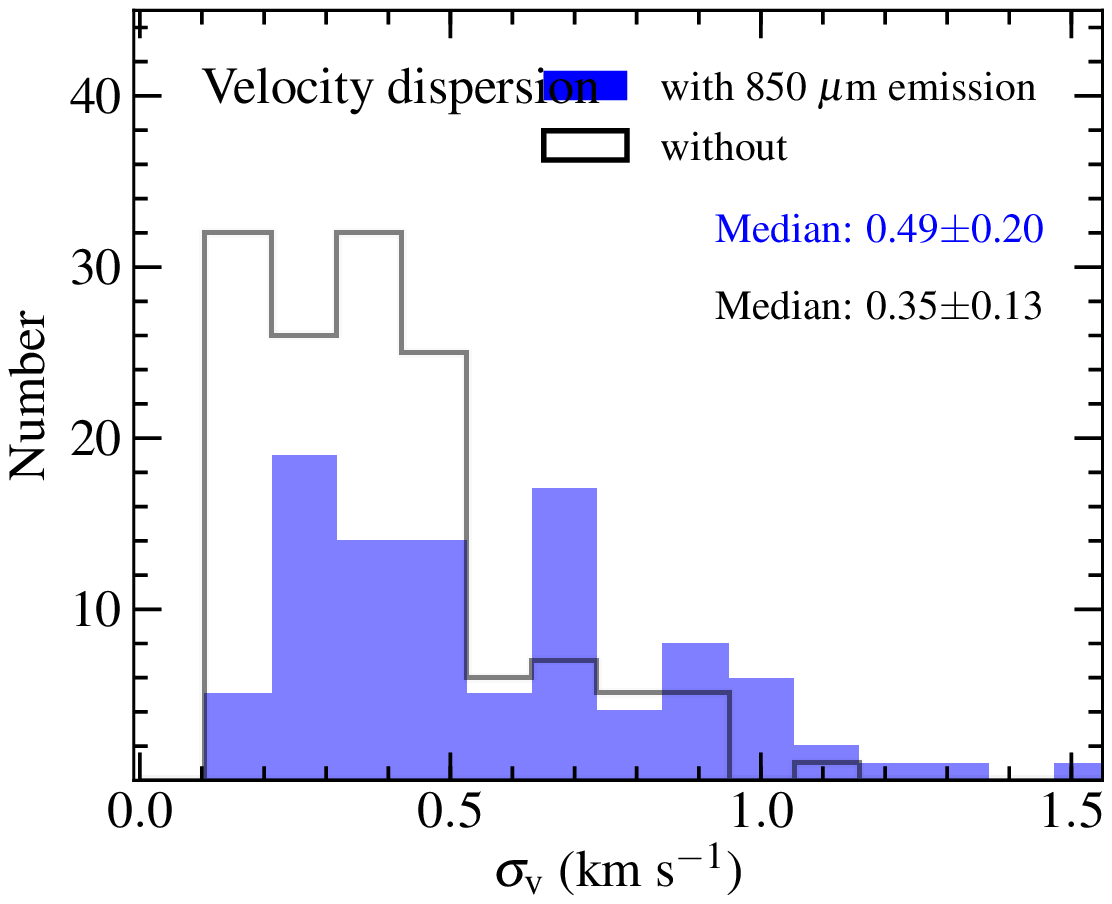}
\includegraphics[width=0.32\textwidth, angle=0]{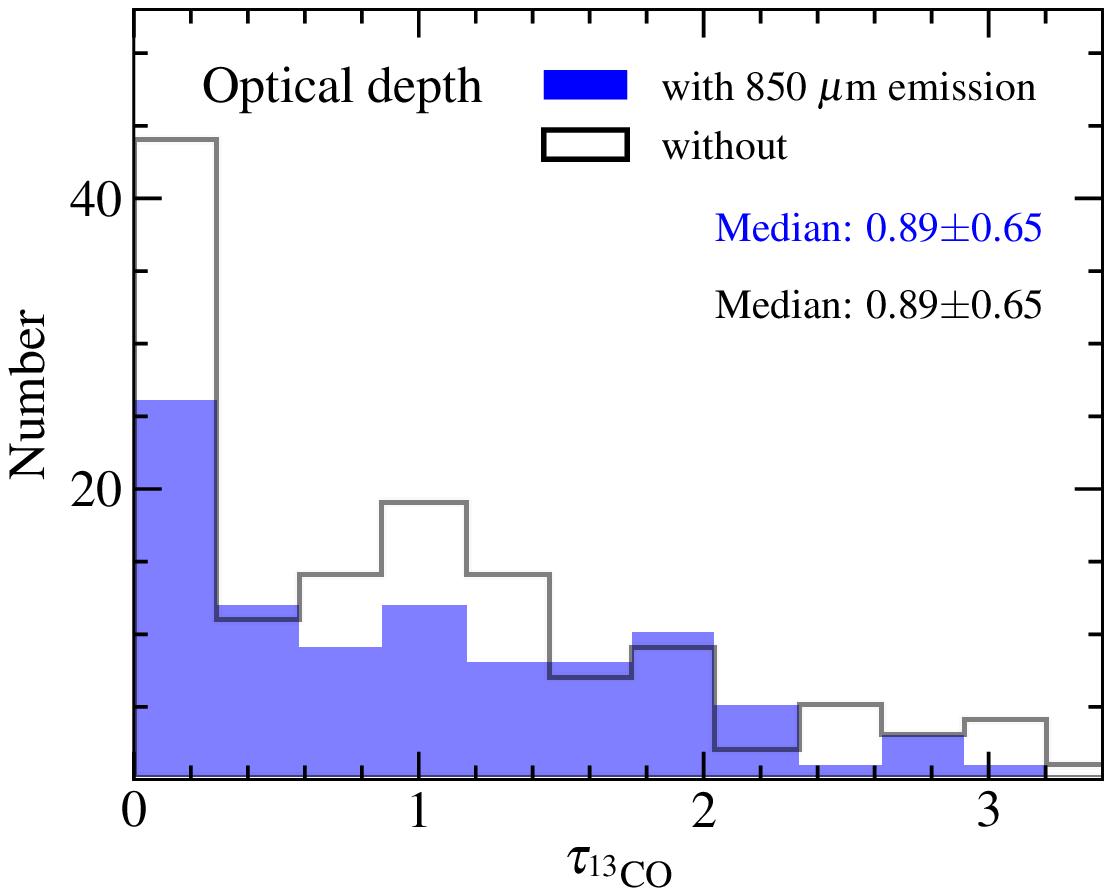}
\includegraphics[width=0.32\textwidth, angle=0]{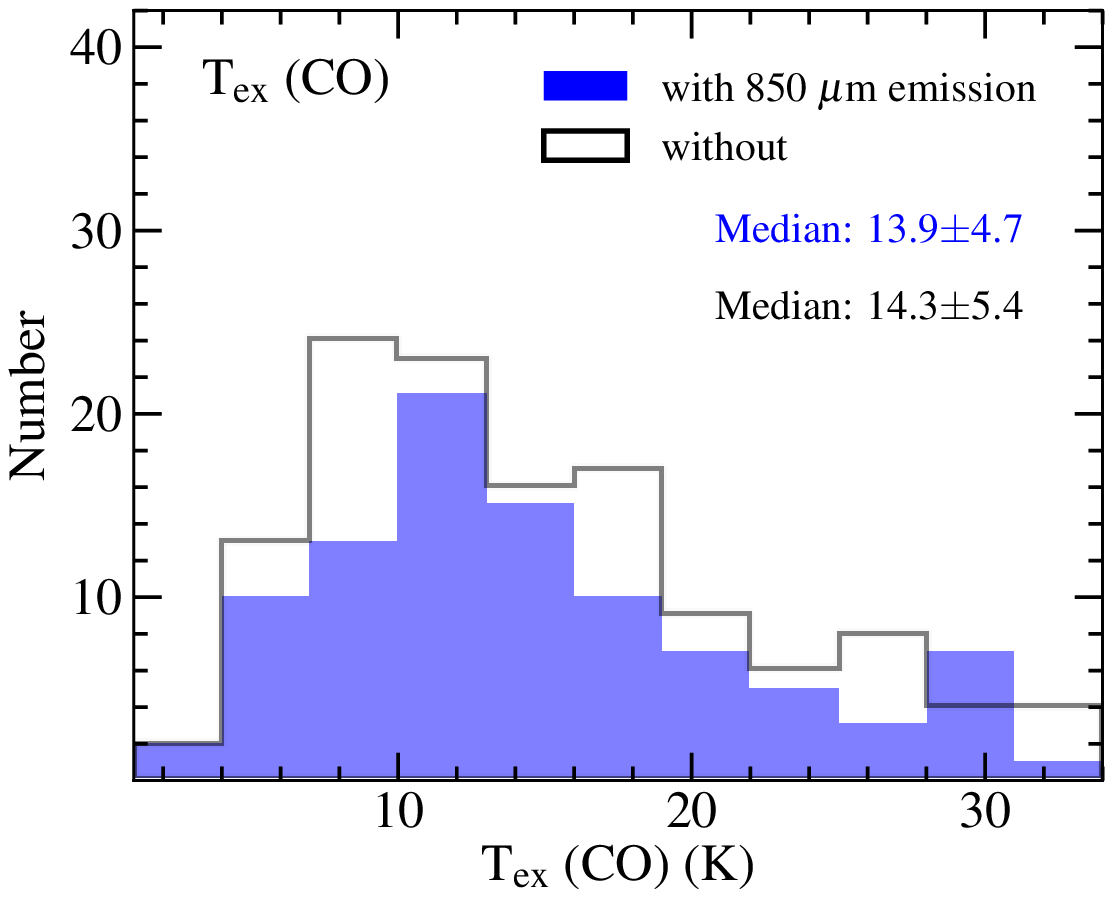}
\includegraphics[width=0.32\textwidth, angle=0]{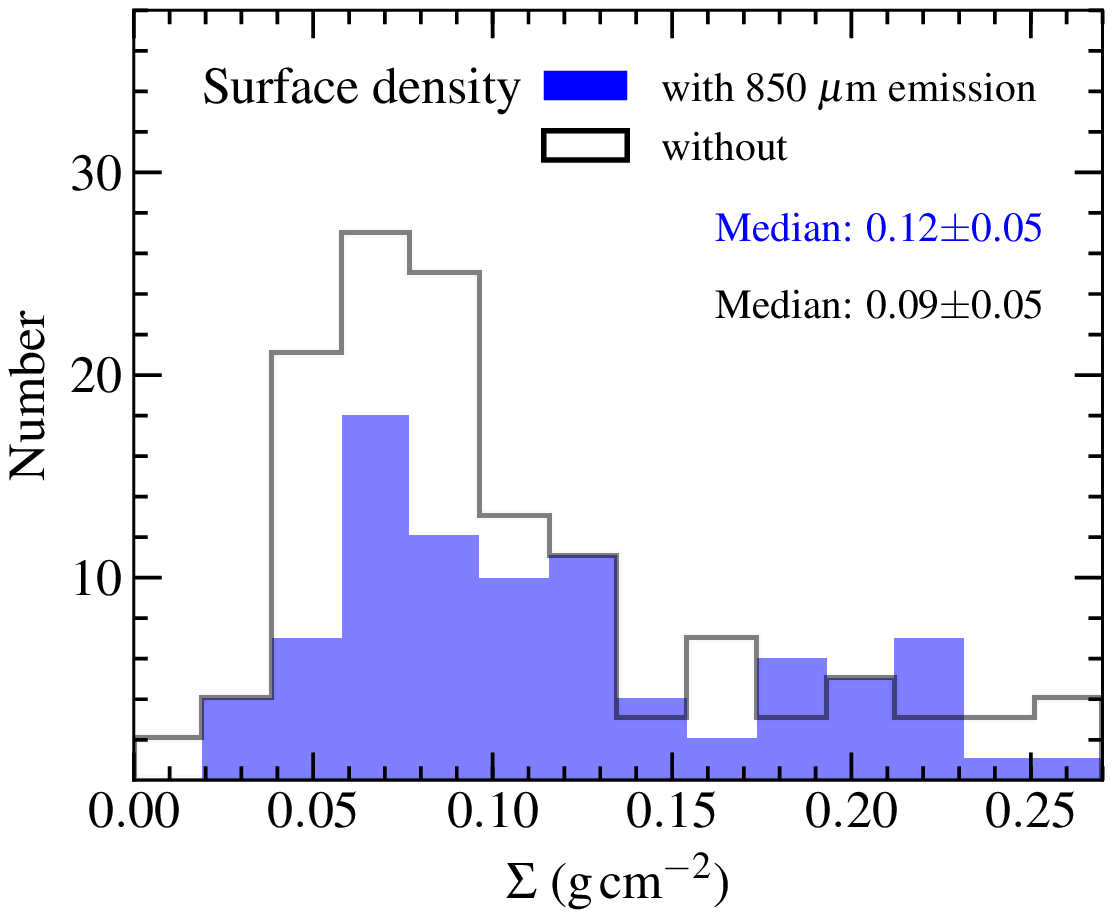}
\includegraphics[width=0.32\textwidth, angle=0]{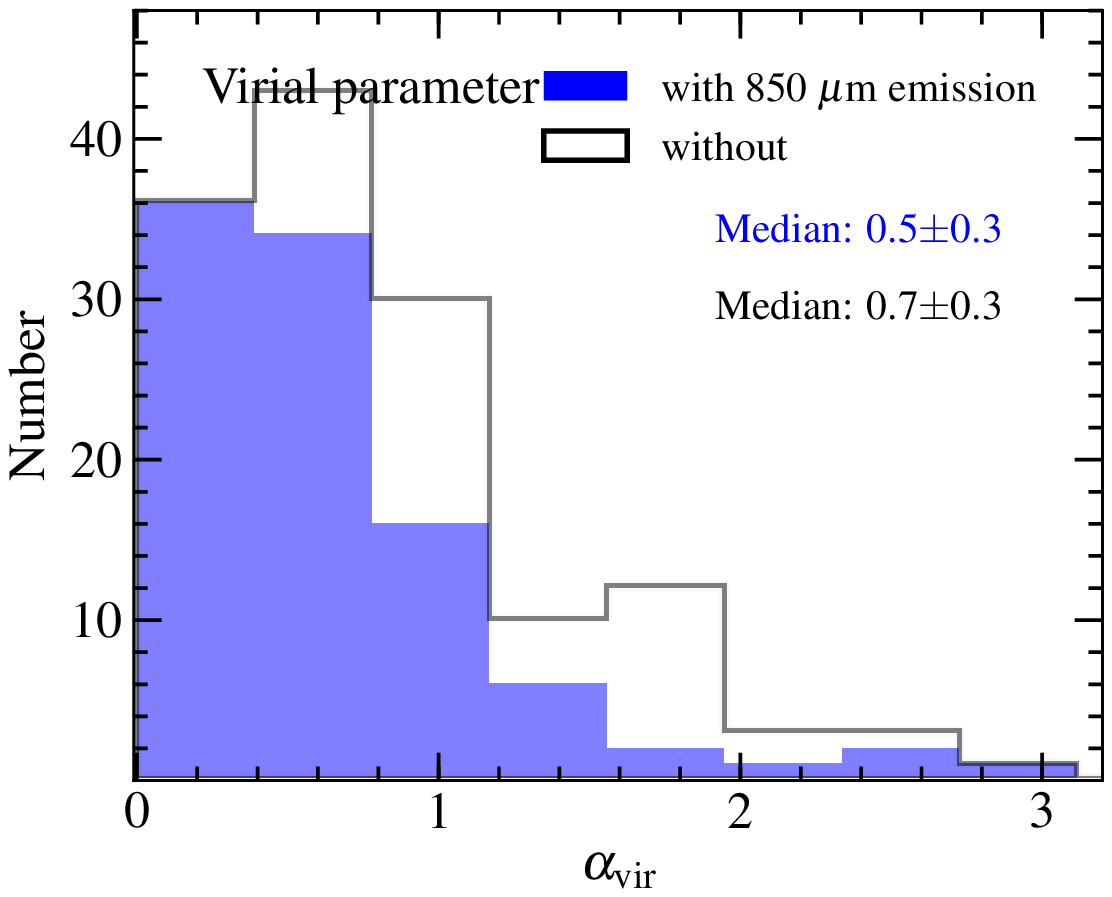}
\caption{Histograms of velocity dispersion, optical depth, excitation temperature, surface density, and virial parameter for the $^{13}$CO clumps in the 64 PGCCs with (\textit{blue}) and without (\textit{black}) 850\,$\mu$m extracted emission. The parameters are listed in Tables \ref{Tab:co_clumps_1}, \ref{Tab:co_clumps_2}, \ref{Tab:850_clumps_1}, and \ref{Tab:850_clumps_2}. The corresponding median value is presented in each sub-histogram. The uncertainty on each median calculated represents median absolute deviation.}
\label{Fig:relations_sep}
\end{figure*}

\subsection{Statistics}

\subsubsection{$^{13}$CO clumps and 850\,$\mu$m cores}

Figure\,\ref{Fig:relations} presents histograms of the velocity dispersions, optical depths, excitation temperatures, surface densities, and virial parameters for the $^{13}$CO clumps and the 850\,$\mu$m cores.

The velocity dispersion ($\sigma_{\rm v}$) histogram in Fig.\,\ref{Fig:relations} shows that the median value is $0.40\pm0.15\,\kms$ for $^{13}$CO clumps, smaller than that ($0.57\pm0.19\, \kms$) of 850\,$\mu$m cores, indicating that the 850\,$\mu$m cores are more dynamically active at a small scale, and being consistent with the fact that 850\,$\mu$m cores are mainly located at the peak positions of $^{13}$CO clumps (see Fig.\,\ref{Fig:13co-850um}), or that some cores with IR emission are forming stars. It seems that the 850\,$\mu$m cores are generally more turbulent than $^{13}$CO clumps. Another possibility is that there is active star formation injecting turbulence in the 850\,$\mu$m cores.

From the optical depth ($\tau_{^{13}\rm CO}$) distribution in Fig.\,\ref{Fig:relations}, we find that the median values are $0.89\pm0.65$ for the $^{13}$CO clumps and $1.75\pm0.46$ for the 850\,$\mu$m cores. Most of the $^{13}$CO clumps have optical depths $< 1.0$. This indicates that most of the $^{13}$CO clumps are more optically thin than the 850\,$\mu$m cores.

The excitation temperature ($T_{\rm ex}$) histogram in Fig.\,\ref{Fig:relations} shows that the median value is $14.1 \pm5.0$\,K for $^{13}$CO clumps, and it is $15.3\pm2.6$\,K for 850\,$\mu$m cores. Considering the 850\,$\mu$m cores are smaller than the $^{13}$CO beam, the filling factors should be $f < 1$. However, we adopt $f = 1$ to estimate excitation temperature, which will lead to underestimate the excitation temperature for 850\,$\mu$m cores (see also error analysis in Section\,\ref{sect:error}). It suggests that the internal parts of the clumps have higher temperatures than the outer parts, probably indicating an internal heating mechanism.

The surface density ($\Sigma$) histogram in Fig.\,\ref{Fig:relations} shows that the median value is $0.10\pm0.04\,\rm g\,cm^{-2}$ for the $^{13}$CO clumps, while it is $0.33\pm0.15\,\rm g\,cm^{-2}$ for the 850\,$\mu$m cores. The median value of surface densities of 850\,$\mu$m cores is much larger than that of $^{13}$CO clumps, indicating that some 850\,$\mu$m cores that are gravitationally bound are denser and represent the precise locations where the stars would form inside the clumps.

The virial parameter ($\alpha_{\rm vir}$) histogram in Fig.\,\ref{Fig:relations} shows that the median value is $0.6\pm0.3$ for the $^{13}$CO clumps, while it is $1.6\pm0.7$ for the 850\,$\mu$m cores. The virial parameters above 2.0 may indicate that the fragments have difficulty in forming stars \citep{Kauffmann2008}, without the help of external pressure. Based on the virial parameters in this work, most of the $^{13}$CO clumps are candidates to form dense cores. Further checking their embedded cores at 850\,$\mu$m, the median value of their virial parameters is around 1.6. Therefore, our cores are mostly gravitationally unbound, and may be dispersing at the core scale, or estimates based on $^{13}$CO overestimate the in-core turbulence. We also note that 76 out of 117 cores at 850\,$\mu$m have WISE counterparts (see Section\,\ref{sect:ir_emission}). It is likely that many cores have already formed stars and may be in the process of expansion.

\subsubsection{PGCCs with and without 850\,$\mu$m emission}

Figure\,\ref{Fig:relations_sep} presents histograms of the velocity dispersions, optical depths, excitation temperatures, surface densities, and virial parameters for the detected ${^{13}\rm CO}$ clumps associated with and without 850\,$\mu$m emission.

The velocity dispersion ($\sigma_{\rm v}$) histogram in Fig.\,\ref{Fig:relations_sep} shows that for PGCCs with and without 850\,$\mu$m emission the median values are $0.49\pm0.20$ and $0.35\pm0.13$\,$\kms$, respectively. This indicates that the PGCCs with 850\,$\mu$m emission are more dynamically active and turbulent than those without 850\,$\mu$m emission.

The optical depth ($\tau_{^{13}\rm CO}$) histogram (using $^{13}$CO as the tracer) in Fig.\,\ref{Fig:relations_sep} shows that the median values are the same ($0.89\pm0.65$), for PGCCs with and without 850\,$\mu$m emission.

The excitation temperature ($T_{\rm ex}$) histogram in Fig.\,\ref{Fig:relations_sep} shows that for PGCCs with and without 850\,$\mu$m emission, the median values are $13.9\pm4.7$ and $14.3\pm5.4$\,K, respectively. Therefore, the two groups are practically are at the same temperature.

The surface density ($\Sigma$) histogram in Fig.\,\ref{Fig:relations_sep} shows that for PGCCs with and without 850\,$\mu$m emission, the median values are $0.12\pm0.05$ and $0.09\pm0.05\,\rm g\,cm^{-2}$. This indicates that the densities are similar for the both.

The virial parameter ($\alpha_{\rm vir}$) histogram in Fig.\,\ref{Fig:relations_sep} shows that for PGCCs with and without 850\,$\mu$m emission the median values are $0.5\pm0.3$ and $0.7\pm0.3$, respectively. Based on the virial parameters in this work, the PGCCs with 850\,$\mu$m emission probably have a slightly greater potential to form stars than those without 850\,$\mu$m emission.

\subsubsection{Comparison with other studies}

Other studies such as those of IRDCs \citep[e.g.][]{Zhangcp2017} found that the linewidth of the C$^{18}$O $J=1-0$ line ranges from around 2.0 to 6.0\,$\kms$ and the volume density from 870\,$\mu$m continuum measurements is greater than $\rm5.0\times10^{4}\,cm^{-3}$, and most cores have virial parameters $\alpha_{\rm vir} < 1.0$. Most ATLASGAL clumps and cores \citep[e.g.,][]{Csengeri2014,Csengeri2017a,Wienen2015,Wienen2017,Koenig2017,Urquhart2018} are also dynamically active, dense, and gravitationally bound, and are high-mass star formation candidates. In this work, however, we find that the 64 PGCCs are dynamically quiescent, optically thin, non-dense, and gravitationally unbound, the typical values of which are $\sigma_{\rm v}<1.5\,\kms$, $\tau_{^{13}\rm CO}<1.0$, $\rm \Sigma<0.3\,cm^{-2}$, and $\alpha_{\rm vir}\gtrsim1.0$. \citet{Wuyf2012}, \citet{Liu2013}, and \citet{Meng2013} detected relatively low column densities, velocity dispersions, and high virial parameters ($\alpha_{\rm vir} > 1.0$) towards other PGCCs with star formation activities. The consistent results further confirm that the PGCCs are mostly quiescent and lack star forming activities, or are most likely at the very initial evolutionary stages of star formation.

\section{Summary}
\label{sect:summary}

To make progress in understanding the early evolution of molecular clouds and dense cores in a wide range of Galactic environments, we carry out an investigation of 64 PGCCs in the second quadrant of the Milky Way, using $^{13}$CO, C$^{18}$O, and 850\,$\mu$m observations. Through the survey, we study their fragmentation and evolution associated with star formation, and show statistical analysis of the extracted $^{13}$CO clumps and 850\,$\mu$m cores.

We present the maps of all $^{13}$CO, C$^{18}$O, and 850\,$\mu$m observations. Using the \textit{Gaussclumps} procedure in GILDAS, we extracted 468 clumps from the $^{13}$CO integrated line intensity maps and 117 cores from the 850\,$\mu$m continuum images. We present all the observational spectra and the derived integrated-intensity maps of $^{13}$CO and C$^{18}$O, compute and list the physical parameters of the lines and the extracted fragments.

Using the Bayesian Distance Calculator \citep{Reid2016}, we derived the distances of all 64 PGCCs in our samples, which are distributed between 0.42 and 5.0\,kpc in the second quadrant of the Milky Way. We find that 60 PGCCs are located in the Local and Perseus arms or the associated interarm region, with 4 PGCCs in the Outer arm.

Fragmentation analysis show that each PGCC fragments into 7.3 clumps on average in $^{13}$CO emission with sizes of around 0.1 -- 3.2\,pc, and each PGCC detected at 850\,$\mu$m fragments into 4.2 cores at 850\,$\mu$m with effective radii of 0.03 -- 0.48\,pc. We suggest that the fragmentation number may be associated with the fragment size, and the relationship between fragmentation number and the fragment size may reflect the nature of clump and core formation efficiency.

We further studied the the properties of the fragments in mass-size plane. We found that in general, the structure follows a relation that is close to $m\sim r^{1.67}$, which is much shallower than what is predicted by \citet{Larson1981}, but is consistent if these objects undergo quasi-isolated gravitational collapse in a turbulent medium \citep{Ligx2017,Zhangcp2017}. At a given scale, the masses of our PGCCs are around 1/10 of that of the typical Galactic massive star-forming regions. This reflects the uniqueness of the PGCC sample: according to \citep{Ligx2017}, the normalization of mass-size relation is determined by the energy dissipation rate of the ambient turbulence. In our sample the mass-size relation can be explained if the turbulence observed in these clumps is 1/3 times (measured in velocity dispersion) the averaged Galactic value.

Statistics indicate that the 850\,$\mu$m cores are more turbulent, more optically thick, and denser than the $^{13}$CO clumps, suggesting that most 850\,$\mu$m cores are better star-forming candidates than the $^{13}$CO clumps. The excitation temperature histogram may suggest that the inner parts of the clumps have higher temperatures than the outer parts, probably indicating an internal heating mechanism. The PGCCs with 850\,$\mu$m emission are more dynamically active, and have more potential to form stars than those without 850\,$\mu$m emission.

Analysis of the clump and core masses, virial parameter, surface density, and mass-size relation suggests that the PGCCs in the second quadrant of the Milky Way have a low core formation efficiency of $\sim$3.0\%, and most are candidates of low-mass star formation. Comparison with previous studies suggests that the PGCCs are mostly quiescent and lack star forming activities, or are most likely at the very initial evolutionary stages of star formation. As evident from the physical parameters, it seems clear that the clumps/cores in this PGCC sample are not able to form high-mass stars.

\acknowledgments

We firstly thank the anonymous referee for prompting many clarifications of this paper. This work is supported by the National Key Basic Research Program of China (973 Program) 2015CB857100, and the National Natural Science Foundation of China through grants 11703040, 11503035 and 11573036. C.-P. Zhang acknowledges support by the China Scholarship Council in Germany as a postdoctoral researcher (No.\,201704910137). Tie Liu is supported by EACOA fellowship. C.W. L. was supported by Basic Science Research Program though the National Research Foundation of Korea (NRF) funded by the Ministry of Education, Science, and Technology (NRF-2016R1A2B4012593). M.J. acknowledges the support of the Academy of Finland Grant No.\,285769. G.-X. Li is supported by the DFG Cluster of Excellence ``Origin and Structure of the Universe''. K.W. is supported by grant WA3628-1/1 of the German Research Foundation (DFG) through the priority program 1573 (``Physics of the Interstellar Medium''). L.V. Toth acknowledges the support of the OTKA grant NN-111016. We are grateful to the staff at the Qinghai Station of PMO for their assistance during the observations. This publication makes use of data products from the Wide-field Infrared Survey Explorer, which is a joint project of the University of California, Los Angeles, and the Jet Propulsion Laboratory/California Institute of Technology, funded by the National Aeronautics and Space Administration. The JCMT is operated by the East Asian Observatory on behalf of The National Astronomical Observatory of Japan, Academia Sinica Institute of Astronomy and Astrophysics, the Korea Astronomy and Space Science Institute, the National Astronomical Observatories of China and the Chinese Academy of Sciences (Grant No.\,XDB09000000), with additional funding support from the Science and Technology Facilities Council of the United Kingdom and participating universities in the United Kingdom and Canada. The SCUBA-2 data mainly taken in this paper were observed under project code M16AL002.

{\it Facilities:} \facility{PMO}, \facility{JCMT}, \facility{WISE}.

\bibliographystyle{apj}
\bibliography{references}

\clearpage

\begin{deluxetable}{lcccc}
\tabletypesize{\footnotesize}
\tablecaption{64 \textit{Planck} Cold Clumps in the Second Quadrant\label{Tab:64_clumps}}
\tablewidth{0pt}
\tablehead{
\colhead{Name} & \colhead{R.A.(J2000)} & \colhead{DEC.(J2000)}  & \colhead{Distance}  & \colhead{Prob.\tablenotemark{a}} \\
 & \colhead{hh:mm:ss} & \colhead{dd:mm:ss}   & \colhead{kpc}
}
\startdata
G098.50-03.24  &  22:05:00.08  &  +51:33:11.69  &$      1.59   (      0.25 )$&      0.76  \\
G108.85-00.80  &  22:58:51.53  &  +58:57:27.09  &$      3.21   (      0.38 )$&      0.56  \\
G110.65+09.65  &  22:28:00.22  &  +69:01:48.10  &$      0.82   (      0.11 )$&      1.00  \\
G112.52+08.38  &  22:52:47.62  &  +68:49:28.31  &$      0.78   (      0.10 )$&      1.00  \\
G112.60+08.53  &  22:52:54.76  &  +68:59:53.90  &$      0.78   (      0.10 )$&      1.00  \\
G115.92+09.46  &  23:24:04.62  &  +71:08:08.69  &$      0.73   (      0.08 )$&      1.00  \\
G116.08-02.38  &  23:56:41.79  &  +59:45:13.19  &$      0.72   (      0.09 )$&      1.00  \\
G116.12+08.98  &  23:28:14.03  &  +70:45:12.38  &$      0.73   (      0.08 )$&      1.00  \\
\enddata
\tablenotetext{a}{The distance probabilities derived by the Bayesian Distance Calculator.}
\end{deluxetable}

\begin{deluxetable}{lccccccccc}
\tabletypesize{\scriptsize}
\rotate
\tablecaption{Observed parameters of extracted CO clumps\label{Tab:co_clumps_1}}
\tablewidth{0pt}
\tablehead{
\colhead{Name} & \colhead{Offset(R.A. DEC.)\tablenotemark{a}} & \colhead{$V_{\rm ^{13}CO}$} & \colhead{$\Delta V_{\rm ^{13}CO}$} & \colhead{$T_{\rm ^{13}CO}$} & \colhead{$V_{\rm C^{18}O}$} & \colhead{$\Delta V_{\rm C^{18}O}$\tablenotemark{b}} & \colhead{$T_{\rm C^{18}O}$}  & \colhead{$\tau_{\rm ^{13}CO}$\tablenotemark{c}} & \colhead{$T_{\rm ex}({\rm ^{13}CO})$\tablenotemark{d}} \\
 & \colhead{($''$ $''$)} & \colhead{$\kms$} & \colhead{$\kms$} & \colhead{K} & \colhead{$\kms$} & \colhead{$\kms$} & \colhead{K}  &  & \colhead{K} }
\startdata
G098.50-03.24$\_$1 &    -42.41,   -118.79 &$       -6.12 \pm        0.01 $&$        1.33 \pm        0.03 $&$        3.09 \pm        0.11 $&$       -5.80 \pm        0.07 $&$        0.93 \pm        0.18 $&$        0.54 \pm        0.10 $&$        0.93 \pm        0.50 $&$ 10.11 \pm 1.80 $ \\
G098.50-03.24$\_$2 &    103.25,     56.95 &$       -5.95 \pm        0.01 $&$        0.81 \pm        0.02 $&$        3.67 \pm        0.12 $&$       -5.91 \pm        0.04 $&$        0.59 \pm        0.12 $&$        0.70 \pm        0.09 $&$        1.19 \pm        0.40 $&$ 11.10 \pm 1.07 $ \\
G108.85-00.80$\_$1 &     21.64,     16.41 &$      -49.53 \pm        0.01 $&$        2.94 \pm        0.03 $&$        5.44 \pm        0.15 $&$      -49.59 \pm        0.08 $&$        2.00 \pm        0.19 $&$        0.91 \pm        0.12 $&$        0.93 \pm        0.38 $&$ 16.93 \pm 2.35 $ \\
G108.85-00.80$\_$2 &    364.31,    146.82 &$      -49.78 \pm        0.02 $&$        2.68 \pm        0.04 $&$        6.02 \pm        0.24 $&$      -49.75 \pm        0.07 $&$        1.52 \pm        0.15 $&$        1.42 \pm        0.20 $&$        1.97 \pm        0.50 $&$ 15.46 \pm 0.85 $ \\
G108.85-00.80$\_$3 &   -160.17,    -58.88 &$      -50.51 \pm        0.02 $&$        2.92 \pm        0.05 $&$        4.16 \pm        0.19 $&$      -50.49 \pm        0.14 $&$        2.45 \pm        0.31 $&$        0.66 \pm        0.15 $&$        0.79 \pm        0.60 $&$ 14.10 \pm 3.94 $ \\
G108.85-00.80$\_$4 &    241.50,    186.31 &$      -49.55 \pm        0.02 $&$        2.31 \pm        0.04 $&$        4.68 \pm        0.16 $&$      -49.74 \pm        0.11 $&$        1.55 \pm        0.20 $&$        0.64 \pm        0.14 $&$        0.41 \pm        0.14 $&$ 21.10 \pm 7.03 $ \\
\enddata
\tablenotetext{a}{The absolute coordinate of each source is listed in Table\,\ref{Tab:64_clumps}.}
\tablenotetext{b}{\bf If the linewidth of C$^{18}$O is less than two velocity channels (or two times velocity resolution), the Gaussian fitting will produce a linewidth of $0.33\pm0.07\,\kms$.}
\tablenotetext{c}{\bf The optical depth with $\tau_{\rm ^{13}CO}<0.20$ is unreliable, therefore we set them as $\tau_{\rm ^{13}CO}=0.20\pm0.05$.}
\tablenotetext{d}{\bf The excitation temperature with $T_{\rm ex}({\rm ^{13}CO})>50.00$\,K is unreliable, therefore we set them as $T_{\rm ex}({\rm ^{13}CO})=50.00\pm5.00$\,K. }
\end{deluxetable}

\begin{deluxetable}{lcccccccccc}
\tabletypesize{\scriptsize}
\rotate
\tablecaption{Derived parameters of extracted CO clumps\label{Tab:co_clumps_2}}
\tablewidth{0pt}
\tablehead{
\colhead{Name} & \colhead{FWHM\tablenotemark{a}} & \colhead{$R_{\rm eff}$\tablenotemark{b}} & \colhead{$\sigma_{\rm C^{18}O}$} & \colhead{$N_{\rm H_{2}}^{\rm ^{13}\rm CO}$} & \colhead{$n_{\rm H_{2}}^{\rm ^{13}\rm CO}$} & \colhead{$\Sigma^{\rm ^{13}\rm CO}$} &  \colhead{$M_{\rm H_{2}}^{\rm ^{13}\rm CO}$} & \colhead{$M_{\rm vir}$}  & \colhead{$\alpha_{\rm vir}$}  \\
 & \colhead{$''$} & \colhead{pc} & \colhead{$\kms$} &  \colhead{$\rm 10^{21} cm^{-2}$} &  \colhead{$\rm 10^3 cm^{-3}$} & \colhead{$\rm g \, cm^{-2}$} & \colhead{$\Msun$} & \colhead{$\Msun$} & }
\startdata
G098.50-03.24$\_$1 &$     162.0 $&$        0.75 $&$        0.39 \pm        0.08 $&$         7.7 \pm         2.0 $&$         4.2 \pm         1.1 $&$       0.089 \pm       0.023 $&$       240.8 \pm        61.4 $&$        87.8 \pm        17.5 $&$        0.36 \pm        0.12 $&$ $ \\
G098.50-03.24$\_$2 &$     128.4 $&$        0.59 $&$        0.25 \pm        0.05 $&$         7.0 \pm         1.3 $&$         4.0 \pm         0.8 $&$       0.068 \pm       0.013 $&$       114.4 \pm        21.7 $&$        44.5 \pm         9.1 $&$        0.39 \pm        0.11 $&$ $ \\
G108.85-00.80$\_$1 &$     166.3 $&$        1.55 $&$        0.85 \pm        0.08 $&$        41.5 \pm         7.0 $&$        10.9 \pm         1.8 $&$       0.488 \pm       0.082 $&$      5630.1 \pm       945.5 $&$       392.1 \pm        36.8 $&$        0.07 \pm        0.01 $&$ $ \\
G108.85-00.80$\_$2 &$     106.8 $&$        1.00 $&$        0.64 \pm        0.06 $&$        68.3 \pm        12.6 $&$        27.5 \pm         5.1 $&$       0.788 \pm       0.145 $&$      3749.7 \pm       690.7 $&$       190.6 \pm        18.8 $&$        0.05 \pm        0.01 $&$ $ \\
G108.85-00.80$\_$3 &$     164.5 $&$        1.53 $&$        1.04 \pm        0.13 $&$        25.5 \pm         7.0 $&$         4.9 \pm         1.4 $&$       0.217 \pm       0.060 $&$      2440.1 \pm       671.5 $&$       474.2 \pm        59.7 $&$        0.19 \pm        0.06 $&$ $ \\
G108.85-00.80$\_$4 &$     170.7 $&$        1.59 $&$        0.66 \pm        0.09 $&$        21.5 \pm         4.3 $&$         4.3 \pm         0.9 $&$       0.196 \pm       0.039 $&$      2379.7 \pm       475.9 $&$       311.5 \pm        40.8 $&$        0.13 \pm        0.03 $&$ $ \\
G110.65+09.65$\_$1 &$     240.7 $&$        0.58 $&$        0.44 \pm        0.02 $&$        40.3 \pm         2.4 $&$        18.4 \pm         1.1 $&$       0.303 \pm       0.018 $&$       479.9 \pm        28.1 $&$        74.9 \pm         3.2 $&$        0.16 \pm        0.01 $&$ $ \\
\enddata
\tablenotetext{a}{The extracted clump sizes with \textit{Gaussclumps} procedure (see Section\,\ref{sect_extraction}).}
\tablenotetext{b}{The clump effective radius $R_{\rm eff} = \rm FWHM/(2\sqrt{ln2})$. The uncertainties of $R_{\rm eff}$ are $\sim$10\%, which will cause additional uncertainty of around 21\% in the derived masses.}
\end{deluxetable}

\begin{deluxetable}{lccccccccc}
\tabletypesize{\scriptsize}
\rotate
\tablecaption{Observed parameters of extracted 850\,$\mu$m cores\label{Tab:850_clumps_1}}
\tablewidth{0pt}
\tablehead{
\colhead{Name} & \colhead{Offset(R.A. DEC.)\tablenotemark{a}} & \colhead{$V_{\rm ^{13}CO}$} & \colhead{$\Delta V_{\rm ^{13}CO}$} & \colhead{$T_{\rm ^{13}CO}$} & \colhead{$V_{\rm C^{18}O}$} & \colhead{$\Delta V_{\rm C^{18}O}$} & \colhead{$T_{\rm C^{18}O}$} & \colhead{$\tau_{\rm ^{13}CO}$} & \colhead{$T_{\rm ex}({\rm ^{13}CO})$} \\
 & \colhead{($''$ $''$)} & \colhead{$\kms$} & \colhead{$\kms$} & \colhead{K} & \colhead{$\kms$} & \colhead{$\kms$} & \colhead{K} &  & \colhead{K} }
\startdata
G108.85-00.80$\_$1 &     78.42,     89.09 &$      -49.46 \pm        0.02 $&$        3.06 \pm        0.05 $&$        6.26 \pm        0.29 $&$      -49.53 \pm        0.10 $&$        2.01 \pm        0.23 $&$        1.22 \pm        0.23 $&$        1.36 \pm        0.57 $&$ 17.12 \pm 1.95 $ \\
G108.85-00.80$\_$2 &    -49.32,    -80.21 &$      -49.99 \pm        0.03 $&$        3.17 \pm        0.07 $&$        4.95 \pm        0.32 $&$      -49.84 \pm        0.16 $&$        2.84 \pm        0.34 $&$        0.94 \pm        0.22 $&$        1.29 \pm        0.70 $&$ 14.25 \pm 2.19 $ \\
G108.85-00.80$\_$3 &     64.49,     46.21 &$      -49.43 \pm        0.02 $&$        2.76 \pm        0.05 $&$        6.44 \pm        0.30 $&$      -49.50 \pm        0.08 $&$        1.79 \pm        0.18 $&$        1.41 \pm        0.21 $&$        1.72 \pm        0.51 $&$ 16.68 \pm 1.22 $ \\
G108.85-00.80$\_$4 &    107.96,    184.14 &$      -49.51 \pm        0.03 $&$        3.14 \pm        0.08 $&$        4.07 \pm        0.29 $&$      - $&$      - $&$      - $&$      - $&$ - $ \\
G108.85-00.80$\_$5 &    -49.38,   -113.27 &$      -49.76 \pm        0.04 $&$        3.22 \pm        0.09 $&$        3.76 \pm        0.32 $&$      -48.72 \pm        0.13 $&$        1.19 \pm        0.33 $&$        1.08 \pm        0.26 $&$        2.71 \pm        1.03 $&$ 9.96 \pm 0.92 $ \\
G108.85-00.80$\_$6 &   -102.54,    -65.87 &$      -51.07 \pm        0.03 $&$        2.37 \pm        0.07 $&$        5.57 \pm        0.36 $&$      - $&$      - $&$      - $&$      - $&$ - $ \\
G108.85-00.80$\_$7 &    -68.94,    -67.74 &$      -50.71 \pm        0.03 $&$        2.80 \pm        0.06 $&$        5.46 \pm        0.28 $&$      -50.72 \pm        0.14 $&$        2.52 \pm        0.27 $&$        1.07 \pm        0.22 $&$        1.37 \pm        0.65 $&$ 15.26 \pm 1.89 $ \\
\enddata
\tablenotetext{a}{The absolute coordinate of each source is listed in Table\,\ref{Tab:64_clumps}.}
\end{deluxetable}

\begin{deluxetable}{lcccccccccc}
\tabletypesize{\scriptsize}
\rotate
\tablecaption{Derived parameters of extracted 850\,$\mu$m cores\label{Tab:850_clumps_2}}
\tablewidth{0pt}
\tablehead{
\colhead{Name} & \colhead{FWHM\tablenotemark{a}} & \colhead{$R_{\rm eff}$\tablenotemark{b}} & \colhead{$\sigma_{\rm C^{18}O}$} & \colhead{$S_{\rm 850}$} & \colhead{$N_{\rm H_{2}}^{\rm 850{\mu}m}$} & \colhead{$n_{\rm H_{2}}^{\rm 850{\mu}m}$} & \colhead{$\Sigma^{\rm 850{\mu}m}$} & \colhead{$M_{\rm H_{2}}^{\rm 850{\mu}m}$} & \colhead{$\alpha_{\rm vir}$} & \colhead{Infrared\tablenotemark{c}} \\
 & \colhead{$''$} & \colhead{pc} & \colhead{$\kms$} &   \colhead{Jy} & \colhead{$\rm 10^{22} cm^{-2}$} & \colhead{$\rm 10^4cm^{-3}$} & \colhead{$\rm g \, cm^{-2}$} & \colhead{$\Msun$}  & & \colhead{WISE}}
\startdata
G108.85-00.80$\_$1 &$      30.2 $&$        0.28 $&$        0.85 \pm        0.10 $&$        2.19 $&$         6.4 \pm         1.5 $&$         5.0 \pm         0.9 $&$       0.403 \pm       0.072 $&$       154.0 \pm        27.6 $&$        0.46 \pm        0.10 $& yes  \\
G108.85-00.80$\_$2 &$      26.5 $&$        0.25 $&$        1.21 \pm        0.14 $&$        1.43 $&$         4.6 \pm         1.4 $&$         6.5 \pm         1.7 $&$       0.464 \pm       0.122 $&$       135.9 \pm        35.7 $&$        0.65 \pm        0.19 $& no  \\
G108.85-00.80$\_$3 &$      30.7 $&$        0.29 $&$        0.76 \pm        0.08 $&$        1.55 $&$         7.0 \pm         1.4 $&$         3.5 \pm         0.4 $&$       0.289 \pm       0.034 $&$       113.8 \pm        13.3 $&$        0.57 \pm        0.09 $& no  \\
G108.85-00.80$\_$4 &$      18.7 $&$        0.17 $&$      - $&$        0.53 $&$        18.8 \pm         0.8 $&$        12.4 \pm         1.6 $&$       0.621 \pm       0.081 $&$        90.5 \pm        11.8 $&$      - $& no  \\
G108.85-00.80$\_$5 &$      16.3 $&$        0.15 $&$        0.51 \pm        0.14 $&$        0.37 $&$         5.3 \pm         1.7 $&$        14.1 \pm         2.7 $&$       0.615 \pm       0.118 $&$        68.2 \pm        13.1 $&$        0.33 \pm        0.11 $& no  \\
G108.85-00.80$\_$6 &$      20.0 $&$        0.19 $&$      - $&$        0.69 $&$        22.5 \pm         0.8 $&$         8.0 \pm         0.7 $&$       0.428 \pm       0.039 $&$        71.0 \pm         6.5 $&$      - $& yes  \\
G108.85-00.80$\_$7 &$      17.6 $&$        0.16 $&$        1.07 \pm        0.11 $&$        0.38 $&$         4.9 \pm         1.4 $&$         5.4 \pm         1.1 $&$       0.253 \pm       0.052 $&$        32.3 \pm         6.6 $&$        1.60 \pm        0.37 $& yes  \\
\enddata
\tablenotetext{a}{The extracted clump sizes with \textit{Gaussclumps} procedure (see Section\,\ref{sect_extraction}).}
\tablenotetext{b}{The clump effective radius $R_{\rm eff} = \rm FWHM/(2\sqrt{ln2})$. The uncertainties of $R_{\rm eff}$ are $\sim$10\%, which will cause additional uncertainty of around 21\% in the derived masses.}
\tablenotetext{c}{The point source cross identification using the AllWISE Data in VizieR Online Data Catalog (see Section\,\ref{sect:obs_wise}).}
\end{deluxetable}




\end{document}